\documentclass[11pt]{article}
\usepackage[T1]{fontenc}
\usepackage{lmodern}
\usepackage{amsmath,amssymb,amsthm,mathtools}
\usepackage[margin=1in]{geometry}
\usepackage[hidelinks]{hyperref}
\usepackage{microtype}
\allowdisplaybreaks
\newtheorem{theorem}{Theorem}[section]
\newtheorem{proposition}[theorem]{Proposition}
\newtheorem{lemma}[theorem]{Lemma}
\newtheorem{corollary}[theorem]{Corollary}
\theoremstyle{remark}
\newtheorem{remark}[theorem]{Remark}
\newcommand{\E}{\mathbb E}
\newcommand{\R}{\mathbb R}
\newcommand{\Z}{\mathbb Z}
\newcommand{\U}{\mathcal U}
\newcommand{\ddk}{\frac{dk}{(2\pi)^d}}
\newcommand{\Var}{\operatorname{Var}}
\newcommand{\tr}{\operatorname{tr}}
\newcommand{\G}{\mathcal G}
\title{Quantitative Gaussian limits of periodic \(O(N)\) models\\
at and below the critical temperature}
\author{Weiqi Jiang\textsuperscript{1}, Pengcheng Hou\textsuperscript{2}, and Zhiyi Li\textsuperscript{2,3}\\[3pt]
\small \textsuperscript{1}Institute of Theoretical Physics, Chinese Academy of Sciences,
Beijing 100190, China\\
\small \textsuperscript{2}Hefei National Laboratory, University of Science and Technology of China,
Hefei 230088, China\\
\small \textsuperscript{3}Department of Modern Physics, University of Science and Technology of China,
Hefei 230026, China\\
\small Corresponding author: Weiqi Jiang, \href{mailto:jiangweiqi@itp.ac.cn}{\nolinkurl{jiangweiqi@itp.ac.cn}}}
\date{}
\hypersetup{pdfauthor={Weiqi Jiang, Pengcheng Hou, Zhiyi Li},
  pdftitle={Quantitative Gaussian limits of periodic O(N) models at and below the critical temperature},
  pdfsubject={Research article}}
\begin{document}
\maketitle
\begin{abstract}
We prove a volume-uniform Gaussian approximation for local spin
coordinates in the zero-field nearest-neighbour $O(N)$ model on periodic
boxes of side $M$ in $d\ge3$, at and below the critical temperature of the
limiting spherical model. On a fixed spatial window, for sufficiently
large $M,N$, the unnormalized $1$-Wasserstein error for the first
$p\le N$ coordinates is at most $C\sqrt p\,(N^{-1/2}+s_M)$, where
$s_M\to0$ is the spectral shift of the finite-volume mean spherical
model. The estimate is uniform on compact ranges of positive directional
couplings and inverse temperatures. The target has independent
coordinate fields, each a scaled anisotropic Gaussian free field with an
independent spatially constant Gaussian term below criticality.

On a window containing a nearest-neighbour star (a site and its $d$
positive-direction neighbours), the error tends to zero as $M,N\to\infty$
exactly when $p=o(N)$ and $\sqrt p\,s_M\to0$. For any choice of periodic
infinite-volume accumulation states, the error on any nonempty fixed window
tends to zero as $N\to\infty$ exactly when $p=o(N)$. In $d\ge5$, when the
inverse temperature exceeds its critical value by at least a fixed positive
amount, the bound improves to $C(p/N+\sqrt p\,M^{-d})$. For fixed $p$, the
sharp order is $N^{-1}+M^{-d}$ on star-containing windows and $N^{-1}$ on
every nonempty fixed window in periodic accumulation states.
\end{abstract}

\section{Introduction}\label{main:introduction}
How large must a periodic box be for the large-$N$ Gaussian
approximation to describe local spin measurements? In the
nearest-neighbour $O(N)$ model the spin length is fixed at $\sqrt N$
and the box has side length $M$. Convergence as $N\to\infty$ at fixed
$M$ gives no error estimate for a box that grows with $N$ or for a
state obtained by taking volume first.

For the isotropic periodic model at zero field, Aru and Korzhenkova
identify the local law obtained by taking $N\to\infty$ before
$M\to\infty$ \cite[Theorems~4.1--4.2]{AK}. At the critical temperature
of the limiting spherical theory, it is a scaled massless Gaussian
free field (GFF). Below that temperature it also contains an
independent Gaussian shift shared by all sites. They ask whether
the limits can be interchanged \cite[Section~1.2]{AK}. See also
\cite[Theorem~5.1.2 and Section~1.5, question~8]{Korzhenkova}.
We control this approximation uniformly in volume at and below the
limiting critical temperature, allowing positive directional
couplings.

For the first $p\le N$ spin coordinates on a fixed spatial window,
Theorem~\ref{main:joint} bounds the unnormalized $1$-Wasserstein
distance to the infinite-lattice Gaussian law for sufficiently large
$M,N$ by
$C_{\mathcal U}\sqrt p\,(N^{-1/2}+s_M)$, uniformly over the compact
parameter ranges in Section~\ref{main:model}. The spectral shift
$s_M$ is determined by the finite-volume spherical constraint
\eqref{main:mass}. For fixed $p$, the approximation error tends to zero along every joint
path. At fixed $\beta,J$, the joint limit is the same Gaussian law obtained
from any choice of periodic accumulation states when volume is taken first
(Corollary~\ref{main:interchange}).

When $p$ grows with $M$ and $N$, two lower bounds limit the
approximation. The one-site sphere marginal detects the fixed spin
length \cite{DF,JM}. On a window containing a site and its $d$
positive-direction neighbours (a nearest-neighbour star), a test
based on spin differences detects the finite-volume error. The
conditions $p=o(N)$ and $\sqrt p\,s_M\to0$ are necessary and
sufficient on such windows and sufficient on every fixed window
(Theorem~\ref{main:finite-criterion}). For periodic accumulation
states the criterion reduces to $p=o(N)$ on any nonempty fixed
window (Corollary~\ref{main:infinite}).

A sharper estimate holds in the ordered regime away from the
critical point. In $d\ge5$, if $\beta-\beta_{c,J}\ge\kappa>0$,
the upper bound improves to
$C_{\mathcal U,\kappa}(p/N+\sqrt p\,M^{-d})$.
For fixed $p$, matching lower bounds give the sharp order
$N^{-1}+M^{-d}$ on star-containing windows and $N^{-1}$ in
periodic accumulation states. Theorem~\ref{main:ordered-rate} also improves the
fixed-$p$ dependence on $N$ in $d=3,4$.

The relation between spin and spherical models is classical
\cite{BerlinKac1952,Stanley1968,MosheZinnJustin2003}.
Kac and Thompson proved agreement of limiting free energies at
noncritical temperatures \cite{KacThompson1971}. Shcherbina's radial
Hamiltonian compares the free energies per site and spin component
with error of order $(\log N)/N$, uniformly in volume and at every
temperature \cite[Theorem~1, Eqs.~(2.3)--(2.4) and Section~3]{Shcherbina}.
Proposition~\ref{prob:pressure} adapts that comparison to obtain
a volume-uniform $O(N^{-1})$ error.

Above the limiting critical temperature, Kupiainen obtained
volume-uniform $1/N$ expansions for the free energy and correlation
functions of the isotropic nearest-neighbour model \cite{Kupiainen}.
Shcherbina stated joint convergence of fixed-order zero-field
correlations in that regime \cite[Section~4, Remark~1]{Shcherbina}.
Ye and Zhu identify the critical and ordered Gaussian fields
as stationary laws of their infinite-volume limiting dynamics
\cite[Theorem~1.5]{YeZhu}. At small inverse temperature,
$0<\kappa<1/(32d)$ in their notation, they bound the $W_2$ error
for fixed-coordinate marginals of the spin laws and prove weak
convergence to the same one-coordinate law in either order of limits
\cite[Theorems~1.6, 5.15--5.16 and Corollary~5.17]{YeZhu}.

In a uniform field, Pearce and Thompson identified the limiting
one-point functions with those of the mean spherical model by
differentiating the limiting free energy. Their two-point comparison
assumes the required interchanges of limits and derivatives
\cite{PearceThompson1977}. In the field $h(1,\dots,1)$,
Shcherbina proved joint convergence of fixed-order coordinate
correlations, with the field removed only after $M,N\to\infty$
\cite[Theorem~2]{Shcherbina}. At low temperature this order of limits
selects a magnetization direction. The periodic zero-field states
considered here retain $O(N)$ invariance.

Following the reflection-positivity approach suggested in
\cite[Section~1.2]{AK}, we combine Gaussian domination with the
pressure comparison to control the local Gram matrix. The
conditional Haar projection turns that estimate into control of
coordinate laws. The orbit representation and the corresponding
Gaussian mixture are classical \cite[Example~2.1 and
Proposition~2.3]{DiaconisEatonLauritzen}. In the ordered regime a sphere
Ward identity and a second-order Gaussian mixture estimate give
the improved rate. Section~\ref{main:proof} outlines the argument.
Appendix~\ref{spec:appendix} controls the finite-volume spectral shift.

\section{Model and main results}\label{main:model}
Let \(d\ge3\), \(J=(J_1,\ldots,J_d)\) with \(J_i>0\), and
\(j=\sum_iJ_i\). On \(\Lambda_M=(\Z/M\Z)^d\), \(M\ge4\), each
\(S_x\in\R^N\) has length \(\sqrt N\). Let \(\sigma_N\) be uniform
probability on that sphere. The Gibbs law is
\begin{equation}\label{main:gibbs}
 \frac1{Z_{M,N,\beta,J}}
 \exp\left\{\beta\sum_{x\in\Lambda_M}\sum_{i=1}^d
 J_i S_x\cdot S_{x+e_i}\right\}\prod_x\sigma_N(dS_x).
\end{equation}
Each unoriented edge is counted once, and the external field is zero.
Define
\[
 (L_{J,M}f)(x)=\sum_iJ_i[2f(x)-f(x+e_i)-f(x-e_i)],
 \quad \lambda_J(k)=2\sum_iJ_i(1-\cos k_i).
\]
Write \(\mathcal K_M=(2\pi/M)\{0,\ldots,M-1\}^d\), with frequencies
modulo \(2\pi\), and \(V=M^d\). Define
\[
 G_J(x,y)=\int_{[-\pi,\pi]^d}
 \frac{e^{ik\cdot(x-y)}}{\lambda_J(k)}\,\ddk,
 \qquad \beta_{c,J}=G_J(0,0).
\]
These integrals are finite for \(d\ge3\). Here \(\beta_{c,J}\) is
the critical inverse temperature of the infinite-volume mean
spherical model. All temperature comparisons below use this
spherical threshold. The transition point of the fixed-$N$ spin model
is a separate quantity.

Let \(G^0_{J,M}\) be the inverse of \(L_{J,M}\) on zero-sum functions,
extended as zero on constants. For a real finitely supported zero-sum
source \(a\), set
\[
 \widehat a(k)=\sum_xa_xe^{ik\cdot x},\quad
 Q_M^0(a)=\frac1V\sum_{k\ne0}\frac{|\widehat a(k)|^2}{\lambda_J(k)},
 \quad Q_\infty(a)=\int\frac{|\widehat a(k)|^2}{\lambda_J(k)}\,\ddk .
\]
For $M$ large enough that the support embeds injectively, we view each
finitely supported source $a$ as a zero-sum source on $\Lambda_M$ by
periodic identification.

Fix a nonempty set of distinct sites
\(\U=\{x_1,\ldots,x_\ell\}\subset\Z^d\).
The law of the first \(p\) spin coordinates at these sites is
\(\nu_{M,N,\beta,J}^{\U,p}\), on \(\R^{\ell\times p}\).
All Wasserstein distances use the unnormalized Frobenius norm (no
division by \(\sqrt p\); see Section~\ref{main:scales}):
\[
 W_1(\mu,\nu)=\inf\{\E\|X-Y\|_F:\mathcal L(X)=\mu,\mathcal L(Y)=\nu\}.
\]
For \(\beta\ge\beta_{c,J}\), let \(\gamma_{\beta,J}^{\U,p}\) have
\(p\) independent centred Gaussian columns with covariance
\begin{equation}\label{main:target}
 \Sigma_{\beta,J}(x,y)=\frac{G_J(x,y)}{\beta}
                    +1-\frac{\beta_{c,J}}{\beta},\qquad x,y\in\U .
\end{equation}
Its diagonal is one. Each column is \(\beta^{-1/2}\) times an
anisotropic GFF plus an independent Gaussian constant of variance
\(1-\beta_{c,J}/\beta\).
The spatially constant term belongs to the limiting ordered-state
covariance.

Uniform statements fix \(d,\U\), \(0<j_0\le j_1<\infty\), and
\(0<b<B<\infty\), and range over
\begin{equation}\label{main:compact}
 J\in[j_0,j_1]^d,\qquad \beta\in[b,B],\qquad \beta\ge\beta_{c,J}.
\end{equation}
We write \(f\asymp g\) for two-sided bounds with constants uniform over \eqref{main:compact}.
Set
\[
 \tau=\beta-\beta_{c,J},\qquad
 a_d(M)=\begin{cases}M,&d=3,\\\log M,&d=4,\\1,&d\ge5,\end{cases}
 \qquad t_d(M)=\sqrt{a_d(M)/V}.
\]
The positive shift \(s_M=s_{M,\beta,J}>0\) solves
\begin{equation}\label{main:mass}
 \frac1V\sum_{k\in\mathcal K_M}\frac1{\lambda_J(k)+s_M}=\beta .
\end{equation}
The left side is strictly decreasing in \(s_M\), diverges as
\(s_M\downarrow0\) and tends to zero as \(s_M\to\infty\), so the solution
exists and is unique. Following common usage for the
spherical model we call \(s_M\) the mass, although it is the square of
a mass in the convention \(L+m^2I\).
The equation fixes an auxiliary finite-volume Gaussian covariance
\(\beta^{-1}(L_{J,M}+s_MI)^{-1}\) with unit diagonal.
This auxiliary Gaussian is the zero-field mean spherical model on
\(\Lambda_M\): the spherical constraint is imposed on the mean squared
spin in the finite box, as in the comparison model of
\cite{Shcherbina}.
The comparison target in the theorems below remains the
infinite-lattice covariance \eqref{main:target} restricted to \(\mathcal U\).
\begin{theorem}[Uniform local joint approximation]\label{main:joint}
There are \(C,M_0<\infty\), depending only on the fixed data above, such
that for every integer \(M\ge M_0\), \(N\ge\max(4,\ell)\), and
\(1\le p\le N\),
\begin{equation}\label{main:bound}
 W_1(\nu_{M,N,\beta,J}^{\U,p},\gamma_{\beta,J}^{\U,p})
 \le C\sqrt p\left[N^{-1/2}+s_M\right].
\end{equation}
Here $M_0$ is chosen so that $\mathcal U$ embeds injectively in the torus.
\end{theorem}

\begin{proposition}[Mass orders]\label{main:mass-orders}
The mass satisfies
\[
 s_M\asymp\frac1{V\{\tau+t_d(M)\}}
\]
uniformly on the parameter ranges of Theorem~\ref{main:joint}. At criticality its order is
\[
 \rho_d(M)=
 \begin{cases}M^{-2},&d=3,\\
 M^{-2}/\sqrt{\log M},&d=4,\\M^{-d/2},&d\ge5.\end{cases}
\]
If \(\tau\ge\kappa>0\), it is bounded by \(C_\kappa M^{-d}\).
\end{proposition}
It is proved in Appendix~\ref{spec:appendix} as
Proposition~\ref{spec:crossover}.

\begin{theorem}[Finite-spin lower bounds and exact coordinate criterion]
\label{main:finite-criterion}
For $\mathcal U_0=\{0,e_1,\ldots,e_d\}$, all integers $M\ge4$,
$N\ge1$ and $1\le p\le N$, under \eqref{main:compact},
\[
 W_1(\nu_{M,N,\beta,J}^{\mathcal U_0,p},
                    \gamma_{\beta,J}^{\mathcal U_0,p})
 \ge c\sqrt p/M^d.
\]
For $N\ge4$, the same distance also satisfies
\[
 W_1(\nu_{M,N,\beta,J}^{\mathcal U_0,p},
                    \gamma_{\beta,J}^{\mathcal U_0,p})
 \ge c\sqrt p\,s_M-C\sqrt{p/N}.
\]
The constants are uniform on the compact parameter ranges.
For the star or any fixed set containing a translated star, along
$M_n,N_n\to\infty$ with parameters satisfying \eqref{main:compact},
\[
 W_1(\nu_{M_n,N_n,\beta_n,J_n}^{\U,p_n},
                 \gamma_{\beta_n,J_n}^{\U,p_n})\to0
 \quad\Longleftrightarrow\quad
 p_n=o(N_n)\ \hbox{and}\ \sqrt{p_n}\,s_{M_n,\beta_n,J_n}\to0.
\]
At criticality the volume conditions are $p=o(M^4)$ for $d=3$,
$p=o(M^4\log M)$ for $d=4$, and $p=o(M^d)$ for $d\ge5$.
Under $\beta-\beta_{c,J}\ge\kappa>0$, the condition is $p=o(M^{2d})$.
Each of these conditions is accompanied by $p=o(N)$.
\end{theorem}

\begin{corollary}[Periodic accumulation states]\label{main:infinite}
For each \(N\), choose any local accumulation state
\(\mu_{N,\beta,J}\) of the periodic laws as \(M\to\infty\), that is, any
local weak subsequential limit of the periodized torus laws.
Uniformly over all such choices and under \eqref{main:compact},
\[
 W_1(\mu_{N,\beta,J}^{\U,p},\gamma_{\beta,J}^{\U,p})
 \le C\sqrt{p/N},\qquad N\ge1,\quad 1\le p\le N.
\]
For any sequence \(1\le p_N\le N\) (and any parameter sequence in
\eqref{main:compact}), the displayed Wasserstein distance tends to zero
as \(N\to\infty\) if and only if \(p_N=o(N)\).
\end{corollary}
\begin{corollary}[Interchange of limits]\label{main:interchange}
Fix \(\U\) and \(p\), and write
\(W_1=W_1(\nu_{M,N,\beta,J}^{\U,p},\gamma_{\beta,J}^{\U,p})\).
Uniformly over \eqref{main:compact}:
\begin{enumerate}
\item[(i)] \(\sup W_1\to0\) as \(\min(M,N)\to\infty\), the supremum
 being over \eqref{main:compact}; in particular \(W_1\to0\) along every
 joint path;
\item[(ii)] \(\lim_{M\to\infty}\limsup_{N\to\infty}W_1=0\);
\item[(iii)] \(\lim_{N\to\infty}\sup
 W_1(\mu_{N,\beta,J}^{\U,p},\gamma_{\beta,J}^{\U,p})=0\), the supremum
 being over all choices of periodic accumulation states.
\end{enumerate}
\end{corollary}
\begin{proof}
For fixed $p$, the bound of Theorem~\ref{main:joint} is uniform in
$M\ge M_0$ and $N\ge\max(4,\ell)$, and $s_M\le C\rho_d(M)\to0$
uniformly by Proposition~\ref{spec:crossover}. This gives (i), and
also (ii), since $\limsup_{N\to\infty}W_1\le C\sqrt p\,s_M$.
Corollary~\ref{main:infinite} gives (iii).
\end{proof}

\begin{remark}[State and observation scope]\label{main:scope}
All spatial windows are fixed, and the periodic accumulation states
retain global \(O(N)\) invariance. Fixed-\(N\) uniqueness, directional
pure states, arbitrary boundary conditions and growing spatial windows
are outside the theorem. For \(\beta<\beta_{c,J}\) see
Section~\ref{main:discussion}.
\end{remark}

\subsection{Coordinate scales}\label{main:scales}
Table~\ref{main:scale-table} collects the mass orders of
Proposition~\ref{main:mass-orders} and the resulting ranges of
Theorem~\ref{main:finite-criterion}.
\begin{table}[ht]
\centering
\small
\setlength{\tabcolsep}{5pt}
\begin{tabular}{c|c|c}
Regime & Mass \(s_M\) & \shortstack{Exact range on\\star-containing windows}\\ \hline
\(d=3,\ \tau=0\) & \(M^{-2}\)
 & \(p=o(\min\{N,M^4\})\)\\
\(d=4,\ \tau=0\) & \(M^{-2}(\log M)^{-1/2}\)
 & \(p=o(\min\{N,M^4\log M\})\)\\
\(d\ge5,\ \tau=0\) & \(M^{-d/2}\)
 & \(p=o(\min\{N,M^d\})\)\\
\(d\ge3,\ \tau\ge\kappa>0\) & \(M^{-d}\)
 & \(p=o(\min\{N,M^{2d}\})\)
\end{tabular}
\caption{Mass orders and necessary and sufficient coordinate ranges for the finite-spin laws
on fixed star-containing windows. All four ranges are sufficient on
any fixed spatial window.}
\label{main:scale-table}
\end{table}

The criterion also covers temperatures varying through the crossover,
where $\tau\asymp t_d(M)$. In that case the volume condition is
$\sqrt p/[V\{\tau+t_d(M)\}]\to0$. Each finite-$(M,N,\beta)$ law is
compared with the target $\gamma_{\beta,J}^{\U,p}$ at the same $\beta$;
the variance of the Gaussian constant tends to zero when $\tau\to0$.
The mass orders and this interpolation are proved in Appendix~\ref{spec:appendix}.

In the unnormalized Frobenius metric, \(W_1\) controls every
\(1\)-Lipschitz function of the whole \(p\)-column block, for example
the Euclidean norm of the observed part of one spin, which feels the
hard length constraint.
Dividing by \(\sqrt p\) would give
\(W_1/\sqrt p\le C(N^{-1/2}+s_M)\), in which \(p\) no longer appears;
for \(p\sim cN\) this normalized distance is \(O(N^{-1/2})\) in
accumulation states, whereas the unnormalized distance stays bounded
below by a positive constant
(Proposition~\ref{prob:coordinate-obstruction}).

\begin{remark}[Auxiliary Gaussian reference]\label{main:gaussian-reference}
Let \(\gamma_M^{\U,p}\) have \(p\) independent centred Gaussian
columns with covariance \(\Sigma_M=\beta^{-1}(L_{J,M}+s_MI)^{-1}\)
restricted to \(\U\). For isotropic couplings
and fixed \(p\), it is also the limit in law of
\(\nu_{M,N,\beta,J}^{\U,p}\) as \(N\to\infty\) at fixed \(M\)
\cite[Theorem~4.1]{AK}. Appendix~\ref{spec:reference} shows that, on
every fixed window \(\U\), for every \(p\ge1\) and every \(M\) large
enough that \(\U\) embeds, with \(C_{\mathcal U}\) independent of
\(p\), \(M\) and the parameters in \eqref{main:compact},
\begin{equation}
 W_1(\gamma_M^{\mathcal U,p},\gamma_{\beta,J}^{\mathcal U,p})
       \le C_{\mathcal U}\sqrt p\,s_M.
 \label{spec:gaussian-upper}
\end{equation}
For the star \(\U_0\) and every \(p\ge1\),
\begin{equation}
 W_1(\gamma_M^{\mathcal U_0,p},\gamma_{\beta,J}^{\mathcal U_0,p})
       \ge\frac{\sqrt p\,s_M}{4\sqrt\pi\,j}.
 \label{spec:gaussian-lower}
\end{equation}
The lower bound also holds on any fixed window containing a
translated star. Thus on such windows the volume term \(\sqrt p\,s_M\) in
Theorem~\ref{main:joint} is of the order of the distance between the
auxiliary reference \(\gamma_M^{\mathcal U,p}\) and the target.
By the triangle inequality, Theorem~\ref{main:joint} and
\eqref{spec:gaussian-upper} give
\(W_1(\nu_{M,N,\beta,J}^{\U,p},\gamma_M^{\U,p})\le
C\sqrt p\,(N^{-1/2}+s_M)\); whether \(s_M\) can be removed is
discussed in Section~\ref{main:discussion}.
\end{remark}

\subsection{Stronger ordered estimates}
The general bound has an $N^{-1/2}$ spin-dimensional contribution
for fixed $p$. At a positive ordered gap this can be improved.
The following theorem keeps the dependence on $p$ explicit.
\begin{theorem}[Stronger ordered rate with explicit coordinate dependence]
\label{main:ordered-rate}
Fix $\kappa>0$. Under \eqref{main:compact} and
$\beta-\beta_{c,J}\ge\kappa$, for sufficiently large integer $M$,
$N\ge\max(4,\ell)$ and $1\le p\le N$,
\[
 W_1(\nu_{M,N,\beta,J}^{\U,p},\gamma_{\beta,J}^{\U,p})
 \le C_{\U,\kappa}
 \left[\min\left\{\sqrt{p/N},\ p\psi_d(N)\right\}
                         +\sqrt p\,M^{-d}\right],
\]
where, with $L_N=\log(\mathrm e N)$,
\[
 \psi_d(N)=\begin{cases}
 (L_N/N)^{3/4},&d=3,\\
 L_N/N,&d=4,\\
 N^{-1},&d\ge5.
 \end{cases}
\]
For every choice of periodic accumulation states, the same bound holds
with the volume term omitted.
The constant is independent of $p,N,M,\beta,J$ within these ranges.
For each fixed $p$, this gives
$C_{\U,p,\kappa}\{\psi_d(N)+M^{-d}\}$.
For $d\ge5$, the bound simplifies to
\[
 C_{\U,\kappa}\left(\frac pN+\sqrt p\,M^{-d}\right).
\]
In these dimensions, for fixed $p$ and a window containing a translated
nearest-neighbour star,
\[
 c_{\U,p,\kappa}(N^{-1}+M^{-d})
 \le W_1(\nu_{M,N,\beta,J}^{\U,p},\gamma_{\beta,J}^{\U,p})
 \le C_{\U,p,\kappa}(N^{-1}+M^{-d}).
\]
For $d\ge5$, on every nonempty fixed window and for every choice of periodic
accumulation states, the fixed-$p$ distance is bounded above and below
by positive constant multiples of $N^{-1}$.
\end{theorem}

\begin{remark}[The one-site lower bound]\label{main:one-site}
The fixed-$p$ lower bound of order $N^{-1}$ in
Theorem~\ref{main:ordered-rate} comes from a one-site norm test that
uses only the uniform sphere marginal of a single spin and holds for
every \(\beta\). So for $d\ge5$ the theorem shows that, apart from the
volume term $M^{-d}$, the interaction does not worsen the one-site
rate $N^{-1}$.
At criticality the test gives only the lower
bound \(W_1\ge \bar\chi_p(\sqrt N/\bar\chi_N-1)=\bar\chi_p\{1/(4N)+O(N^{-2})\}\), of order
\(\sqrt p/N\), where \(\bar\chi_q\) is the mean norm of a standard
\(q\)-dimensional Gaussian vector; the critical spin-dimensional rate
\(N^{-1/2}\) of Theorem~\ref{main:joint} is therefore not known to be
sharp.
\end{remark}
\section{Outline of the proofs}\label{main:proof}
The pressure derivative is the \(J\)-weighted sum of the nearest-neighbour
spin correlations per component. Gaussian domination bounds the
single-coordinate variances of spin differences by their massless
Gaussian values. Let \(\delta_{M,N}\) be the \(J\)-weighted sum of the
nearest-neighbour variance deficits. The corresponding deficit of the
finite-volume mean spherical model is \(s_M-1/(\beta V)\). Because the
spin and comparison laws both have unit site variances, the difference
between the deficits is determined by the difference of the pressure
derivatives. This accounts for the spectral shift in the error.

Section~\ref{main:gdpressure} compares the two pressures with error
\(O(N^{-1})\), uniformly in \(M\). An upper secant over an interval of
length \(h\) bounds the derivative difference from above. The pressure
comparison contributes an error of order \(1/(Nh)\) and the curvature
of the comparison pressure contributes an error of order \(h\).
Taking \(h=N^{-1/2}\) gives
\begin{equation}\label{main:saturation}
 \delta_{M,N}\le C N^{-1/2}+s_M.
\end{equation}

The variance deficit for a general zero-sum source is a positive
semidefinite quadratic form. Writing a spin difference as a sum along a
nearest-neighbour path extends the bound to each pair in the window.
Gaussian domination also controls the upper tail of the squared norm
of each difference.
The spin length fixes the diagonal of the local Gram matrix, so the
Green-function comparison is needed only for dipoles. Their zero sum
gives a quadrature error of \(O(M^{-d})\), which can be absorbed into
\(s_M\) (Appendix~\ref{spec:appendix}).

By rotation invariance the observed block has the law of \(A^{1/2}Q_N\),
where \(A\) is the local Gram matrix and \(Q_N\) is an independent,
scaled block of a Haar orthogonal matrix. Replacing \(Q_N\) by a
standard Gaussian matrix gives the comparison even for singular \(A\).
The pressure secant, squared-norm fluctuations and Haar approximation
all contribute to the \(O(\sqrt{p/N})\) error.

For the lower bounds, the test compares the norm of the observed
coordinates of a nearest-neighbour difference with the norm of those
coordinates at a single site. The single-site term calibrates the test
and its Lipschitz constant stays bounded as \(p\) grows. The \(J\)-weighted
difference between the target and spin variances is
\(\delta_{M,N}+1/(\beta V)\), so it is at least \(1/(\beta V)\).
A left pressure secant also bounds it below by \(s_M-C/\sqrt N\)
for \(N\ge4\), even if its left endpoint lies below the critical inverse
temperature.
The direction witnessing the weighted shortfall can depend on the
parameters. A star contains every direction that may be needed
(Section~\ref{main:lower}).

When \(\beta-\beta_{c,J}\ge\kappa>0\), a sphere Ward identity sharpens the
deficit estimate. The covariance expansion cancels a term equal to the
deficit times the excess mean squared magnetization above its
finite-volume massless reference value. A one-sided fluctuation
estimate and spectral interpolation then give the improved deficit bound
(Proposition~\ref{ordered:deficit}).

The other gain comes from expanding Gaussian expectations around the
mean local Gram matrix. When this mean is bounded away from
singularity, the linear term averages to zero and the mixture error
is bounded by \(Cp\) times the second moment of the Gram fluctuation.
When \(N<2(V+1)\), the sharper deficit estimate controls these
fluctuations. When \(N\ge2(V+1)\), the full Gram density controls
the variance and a second Ward identity controls the covariance bias.
The split keeps the volume term at \(\sqrt p/V\). Using the deficit
estimate alone would leave a \(p/V\) term when \(N\) is much larger
than \(V\) (Section~\ref{main:ordered}).

\section{Gaussian domination and the pressure comparison}\label{main:gdpressure}
Proposition~\ref{main:gd} gives Gaussian domination for every
\(M\ge4\), including odd \(M\); Proposition~\ref{prob:pressure}
compares the pressure with the Gaussian comparison pressure; and
Lemma~\ref{prob:energy-deficit} turns this comparison into the deficit
bound \eqref{prob:uniform-deficit}.

\subsection{Gaussian domination}\label{main:gdsection}
The following directional-coupling form of the transfer-operator
argument of Fr\"ohlich, Simon and Spencer
\cite[pp.~83--84]{FSS} applies to every integer \(M\ge4\).
\begin{proposition}[Vector-source bound]\label{main:gd}
Let \(\beta>0\), and let \(M\ge4\), even or odd, and \(N\ge1\) be
integers. For \(b_x\in\R^N\) with \(\sum_xb_x=0\), the Gibbs law
\eqref{main:gibbs} satisfies
\begin{equation}\label{main:gdvector}
 \E\exp\left\{\sum_xb_x\cdot S_x\right\}
 \le\exp\left\{\frac1{2\beta}
            \sum_{\alpha=1}^N(b^\alpha,G^0_{J,M}b^\alpha)\right\}.
\end{equation}
In particular, for a real zero-sum source \(a\) and \(v\in\R^N\),
\begin{equation}\label{main:rankone}
 \E\exp\left\{v\cdot\sum_xa_xS_x\right\}
                  \le\exp\{|v|^2Q_M^0(a)/(2\beta)\}.
\end{equation}
\end{proposition}
\begin{proof}[Proof of Proposition~\ref{main:gd}]
For arbitrary bond vectors \(h_{x,i}\), define
\[
 {\cal Z}(h)=\int\exp\left\{-\frac\beta2
 \sum_{x,i}J_i|S_x-S_{x+e_i}-h_{x,i}|^2\right\}
                   \prod_x\sigma_N(dS_x).
\]
We first show \({\cal Z}(h)\le{\cal Z}(0)\).
Temporarily replace the prior by the same smooth nonnegative compactly
supported probability density at each site. Fix direction \(q\).
A transverse layer has variable \(u\in\R^{NM^{d-1}}\).
Let \(F_i\) denote multiplication by the product of the layer priors
and transverse bond factors; this product is a bounded, nonnegative,
integrable function.
Collect the source between layers \(i\) and \(i+1\) into \(a_i\).
On \(L^2(\R^{NM^{d-1}})\), set
\[
 T_a(u,v)=e^{-\beta J_q|u-v-a|^2/2},\qquad U_af(u)=f(u-a).
\]
Gaussian convolution \(T_0\) is bounded and positive by its Fourier
multiplier; translation \(U_a\) is unitary and commutes with \(T_0\)
and its square root. Thus \(T_a=T_0^{1/2}U_aT_0^{1/2}\).
The positive operators \(B_i=T_0^{1/2}F_iT_0^{1/2}\) are trace class:
\(F_i^{1/2}T_0^{1/2}\) is Hilbert--Schmidt, since \(F_i\) is integrable
and the Gaussian half-convolution kernel is square integrable.
Cyclicity and the integral kernels give
\[
 {\cal Z}(h)=\operatorname{Tr}\left(\prod_{i=1}^M B_iU_{a_i}\right)
 \le\prod_{i=1}^M\|B_iU_{a_i}\|_{{\cal S}_M}
 =\prod_{i=1}^M[\operatorname{Tr}(B_i^M)]^{1/M},
\]
by Schatten H\"older with exponent \(M\).
Each trace is the partition function obtained by setting all
direction-\(q\) sources to zero and copying layer \(i\)'s transverse
sources to all layers. Apply this operation in directions \(1,\ldots,d\).
Copying never reintroduces a source in a direction already set to zero.
Every final branch has source zero, and the geometric weights sum to
one. This proves \({\cal Z}(h)\le{\cal Z}(0)\).
The argument uses homogeneous directional couplings and identical priors;
the weights \(J_i\) may differ and \(M\) may be odd.

Choose smooth priors supported in one compact set and weakly converging
to the sphere probability. At fixed \(M,N,h\), the continuous integrand
is bounded there, so product weak convergence restores the sphere
inequality. Because the spin length is fixed, the law normalized by
\({\cal Z}(0)\) coincides with the Gibbs law \eqref{main:gibbs}.
Finally put \(f=\beta^{-1}G^0_{J,M}b\) coordinatewise and
\(h_{x,i}=f_x-f_{x+e_i}\). Summation by parts gives
\[
 \beta\sum_{x,i}J_i h_{x,i}\cdot(S_x-S_{x+e_i})
                 =\sum_xb_x\cdot S_x,\qquad
 \frac\beta2\sum_{x,i}J_i|h_{x,i}|^2
     =\frac1{2\beta}\sum_\alpha(b^\alpha,G^0_{J,M}b^\alpha).
\]
Expanding the shifted partition proves \eqref{main:gdvector};
take \(b_x=a_xv\) to obtain \eqref{main:rankone}.
\end{proof}

\subsection{Uniform pressure comparison}
Write \(P_{M,N}=(NV)^{-1}\log Z_{M,N,\beta,J}\), and define the
Gaussian comparison pressure by
\begin{equation}\label{main:comparison-pressure}
 P_M^{\rm comp}(\beta)
 =\beta j+\frac{\beta s_M}{2}-\frac12-\frac12\log\beta
       -\frac1{2V}\log\det(L_{J,M}+s_MI).
\end{equation}
For this comparison, \eqref{main:mass} defines $s_M$ for every
$\beta>0$. The next proposition is therefore available on both sides
of the limiting critical inverse temperature, as needed for the
pressure secants.
The sphere prior is normalized to a probability measure. Keeping this
normalization is essential: the radial $(\log N)/N$ terms in the
proof cancel, leaving an $O(N^{-1})$ pressure error.
The construction below follows Shcherbina's auxiliary radial
Hamiltonian \cite[Section~3]{Shcherbina}; tracking the exact polar
Jacobian in this way improves her bound of order $(\log N)/N$ to
$O(N^{-1})$ with explicit constants.

\begin{proposition}[Uniform pressure comparison]
\label{prob:pressure}
For every $B'>0$, all $0<\beta\le B'$, all integers $M\ge4$, and all
$N\ge4$,
\begin{equation}
 |P_{M,N}(\beta)-P_M^{\mathrm{comp}}(\beta)|
 \le\varepsilon_N(B',J)
 :=\frac{8(1+4B'j)+3/2}{N}.
 \label{main:pressure-error}
\end{equation}
In particular this estimate is uniform in volume, and is uniform for
$J$ in any positive compact set.
\end{proposition}

\begin{proof}
Put $\mathcal H=L_{J,M}+s_MI$, $z=2j+s_M$,
$\Delta=s_M+4j$, and $t=\beta\Delta$.
The absolute row sum of $\mathcal H$ is $\Delta$.
Since $0\le\lambda_J\le4j$, the mass equation gives
\begin{equation}
 1\le t\le T:=1+4B'j.
 \label{prob:radial-t}
\end{equation}
For $X_x=\rho_xS_x$, with $\rho_x>0$ and $|S_x|^2=N$, the exact
polar Jacobian relative to the probability prior is
\begin{equation}
 dX_x=c_N\rho_x^{N-1}\,d\rho_x\,\sigma_N(dS_x),
 \qquad c_N=\frac{2\pi^{N/2}N^{N/2}}{\Gamma(N/2)}.
 \label{prob:polar-jacobian}
\end{equation}
The Gaussian Hamiltonian and partition function are
\[
 H_G(X)=\frac12\sum_{x,y}\mathcal H_{xy}X_x\cdot X_y,
 \qquad
 Z_G=(2\pi/\beta)^{NV/2}\det(\mathcal H)^{-N/2}.
\]
Under its normalized law the $N$ coordinate fields are independent,
with covariance $\Sigma^G=\beta^{-1}\mathcal H^{-1}$; its restriction to a
window $\mathcal U$ is the covariance $\Sigma_M$ of
Remark~\ref{main:gaussian-reference}.
Translation invariance and \eqref{main:mass} imply $\Sigma^G_{xx}=1$,
hence $|\Sigma^G_{xy}|\le1$.

For angular configurations set
\[
 q_x=\frac1N\sum_y\mathcal H_{xy}S_x\cdot S_y,
 \quad |q_x|\le\Delta,
\]
\[
 I_N(q)=\int_0^\infty\rho^{N-1}
 e^{-\beta N[q(\rho-1)+\Delta(\rho-1)^2]}\,d\rho,
 \quad \Phi_N(q)=\beta^{-1}\log I_N(q).
\]
Consider the auxiliary Hamiltonian
\[
 \Gamma(X)=H_G(S)+N\sum_xq_x(\rho_x-1)
       +\Delta N\sum_x(\rho_x-1)^2+\sum_x\Phi_N(q_x).
\]
The radial integrals cancel exactly, so, with
$Z_A=\int e^{-\beta H_G(S)}\prod_x\sigma_N(dS_x)$,
\begin{equation}
 Z_\Gamma=c_N^VZ_A,
 \qquad Z_{M,N}=e^{\beta zNV/2}Z_A.
 \label{prob:radial-cancellation}
\end{equation}

We retain both scalar prefactors.  Write $\xi=\rho-1$ and
$c=1-1/N-\beta q$.
For $|\xi|\le N^{-1/2}\le1/2$,
$\log(1+\xi)\ge \xi-\xi^2$, and consequently the exponent in $I_N(q)$
is at least $Nc\xi-N(1+t)\xi^2$.
Integration over this symmetric interval and
$\cosh(Nc\xi)\ge1$ give
\begin{equation}
 \log I_N(q)\ge-\frac12\log N+\log2-1-t.
 \label{prob:radial-lower}
\end{equation}
The inequality $\log\rho\le\rho-1$, completion of the square, and
extension of the integral to the whole real line give
\begin{equation}
 \log I_N(q)\le\frac{N}{4t}(1-1/N-\beta q)^2
                 +\frac12\log\frac\pi{tN}.
 \label{prob:radial-upper}
\end{equation}
These inequalities hold for every real $q$.

We next bound the Gaussian expectation of
$\mathcal R=\Gamma-H_G$ using only the diagonal covariance normalization.
Since $N\rho_x^2$ has a chi-square distribution with $N$ degrees of
freedom,
\[
 \mathbb E_G\rho_x^2=1,
 \qquad \mathbb E_G(\rho_x^2-1)^2=2/N,
 \qquad \mathbb E_G(\rho_x-1)^2\le2/N.
\]
Wick's formula gives
$\mathbb E_G\rho_x^2\rho_y^2=1+2(\Sigma^G_{xy})^2/N\le1+2/N$.
Also
\[
 \mathbb E_G\rho_x\rho_y
 =1-\tfrac12\mathbb E_G(\rho_x-\rho_y)^2\ge1-4/N,
 \qquad
 \mathbb E_G(\rho_x\rho_y-1)^2\le10/N.
\]
Put $\zeta_x=N^{-1}X_x\cdot(\mathcal HX)_x$.
Using $\mathcal H\Sigma^G=\beta^{-1}I$, independent coordinate summation
and Wick's formula give
\[
 \mathbb E_G(\beta\zeta_x)=1,
 \qquad
 \operatorname{Var}_G(\beta\zeta_x)
 =\frac{\beta\mathcal H_{xx}+1}{N}\le\frac{t+1}{N}.
\]
The identity
\[
 q_x-\zeta_x=\sum_y\mathcal H_{xy}
 \frac{S_x\cdot S_y}{N}(1-\rho_x\rho_y)
\]
and weighted Cauchy--Schwarz therefore imply
\[
 \mathbb E_G\{\beta^2(q_x-\zeta_x)^2\}\le10t^2/N,
 \qquad
 \mathbb E_G(1-1/N-\beta q_x)^2
 \le\frac{2t+4+20t^2}{N}.
\]
For the second bound one applies $(u+v)^2\le2u^2+2v^2$ to
$u=1-1/N-\beta\zeta_x$ and $v=\beta(\zeta_x-q_x)$.

Expansion gives
\[
 \mathcal R=\Delta N\sum_x(\rho_x-1)^2
 -\frac12\sum_{x,y}\mathcal H_{xy}(\rho_x-1)(\rho_y-1)
                  S_x\cdot S_y+\sum_x\Phi_N(q_x).
\]
The absolute value of the cross term is at most
$\Delta N\sum_x(\rho_x-1)^2/2$.
Thus $\mathcal R\ge\sum_x\Phi_N(q_x)$ pointwise and, by
\eqref{prob:radial-upper},
\[
 \frac\beta V\mathbb E_G\mathcal R
 \le3t+\frac{2t+4+20t^2}{4t}+\frac12\log\frac\pi{tN}
 =8t+\frac12+\frac1t+\frac12\log\frac\pi{tN}.
\]
The Gibbs variational inequalities are
$-\beta\mathbb E_G\mathcal R\le\log(Z_\Gamma/Z_G)
 \le-\beta\mathbb E_\Gamma\mathcal R$.
The lower scalar bound controls the latter expectation; hence
\begin{equation}
 \frac{\frac12\log N-8t-\frac12-\frac1t
                    -\frac12\log(\pi/t)}{N}
 \le\frac1{NV}\log\frac{Z_\Gamma}{Z_G}
 \le\frac{\frac12\log N+1+t-\log2}{N}.
 \label{prob:variational-bounds}
\end{equation}
All the expectations exist: the angular variables range over a compact
set and the auxiliary radial Hamiltonian has a positive quadratic
coefficient.

Stirling's inequalities give
\[
 \log c_N=\frac N2\log(2\pi\mathrm e)+\frac12\log(N/\pi)-\theta_N,
 \qquad 0<\theta_N<\frac1{6N}.
\]
Combining this with \eqref{prob:radial-cancellation} yields the exact
normalization identity
\begin{equation}
 P_{M,N}-P_M^{\mathrm{comp}}
 =\frac1{NV}\log\frac{Z_\Gamma}{Z_G}
   -\frac{\log(N/\pi)}{2N}+\frac{\theta_N}{N}.
 \label{prob:normalization-identity}
\end{equation}
The $\log N$ terms cancel on both sides of
\eqref{prob:variational-bounds}.  The remaining lower bound is
\[
 \frac{-8t-1/2-1/t+(\log t)/2+\theta_N}{N}
 \ge-\frac{8T+3/2}{N}.
\]
The upper bound is at most $(T+1)/N$, because
$\theta_N<1/24<\log2-\tfrac12\log\pi$ for $N\ge4$.
This proves \eqref{main:pressure-error}.
\end{proof}

\subsection{Pressure derivatives and the energy deficit}
By Proposition~\ref{main:gd}, the Gibbs measure satisfies the
hypothesis \eqref{prob:GD} of the next lemma.
\begin{lemma}[Pressure derivatives and finite-volume deficit]
\label{prob:energy-deficit}
Suppose the Gibbs measure satisfies the full-vector Gaussian domination
\begin{equation}
 \mathbb E\exp\!\left\{v\cdot\sum_xa_xS_x\right\}
 \le\exp\!\left\{\frac12K_M(a)|v|^2\right\},
 \quad K_M(a)=\beta^{-1}Q_M^0(a),
 \quad Q_M^0(a)=(a,G^0_{J,M}a),
 \label{prob:GD}
\end{equation}
for every zero-sum real source $a$ and every $v\in\mathbb R^N$.
Let
\[
 D(a,a)=K_M(a)-\mathbb E\frac{|\sum_xa_xS_x|^2}{N},
 \quad b_i=\delta_0-\delta_{e_i},
 \quad \delta_{M,N}=\sum_iJ_iD(b_i,b_i).
\]
Then
\begin{align}
 (P_M^{\mathrm{comp}})'&=j+s_M/2-1/(2\beta),&
 0\le(P_M^{\mathrm{comp}})''&\le1/(2\beta^2),
 \label{prob:pressure-derivatives}\\*
 \delta_{M,N}
 &=2[P'_{M,N}-(P_M^{\mathrm{comp}})']+s_M-\frac1{\beta V}.
 \label{prob:energy-identity}
\end{align}
For $0<h\le B'-\beta$ and $N\ge4$,
\begin{equation}
 0\le\delta_{M,N}
 \le\frac{4\varepsilon_N(B',J)}h+\frac h{2\beta^2}
       +s_M-\frac1{\beta V}.
 \label{prob:secant}
\end{equation}
In particular, on $J\in[j_0,j_1]^d$ and $\beta\in[b,B]$, with
$0<j_0\le j_1$ and $0<b\le B<\infty$, one may take
\begin{equation}
 \delta_{M,N}\le c_*N^{-1/2}+s_M,
 \quad
 c_*=2\sqrt2\max(1,b^{-1})\sqrt{K_P^+},
 \quad K_P^+=8(1+4(B+1)dj_1)+3/2.
 \label{prob:uniform-deficit}
\end{equation}
These assertions hold also for $\beta<\beta_{c,J}$.
\end{lemma}

\begin{proof}
The partial derivative of \eqref{main:comparison-pressure} with respect to its
mass is zero by \eqref{main:mass}.
Moreover
\[
 s_M'(\beta)=-\left[\frac1V\sum_k
              (\lambda_J(k)+s_M)^{-2}\right]^{-1}.
\]
The bracket is at least $\beta^2$ by Jensen, proving
\eqref{prob:pressure-derivatives}.
Global $O(N)$ symmetry and the second derivative at $v=0$ in
\eqref{prob:GD} give $D(a,a)\ge0$.  The Fourier identity
$\sum_iJ_i|\widehat b_i(k)|^2=\lambda_J(k)$ gives
\[
 \sum_iJ_iQ_M^0(b_i)=1-V^{-1}.
\]
Writing $u_{M,N}=P'_{M,N}=\sum_iJ_i
\mathbb E(S_0\cdot S_{e_i})/N$ and using $|S_x|^2=N$, we obtain
$\delta_{M,N}=(1-V^{-1})/\beta-2(j-u_{M,N})$.
This is \eqref{prob:energy-identity}.
Convexity of $P_{M,N}$, the pressure error, and the curvature bound give
\[
 2[P'_{M,N}-(P_M^{\mathrm{comp}})']
 \le\frac{4\varepsilon_N}h+\frac h{2\beta^2},
\]
which proves \eqref{prob:secant}.
For the uniform estimate use $B'=B+1$ and $\varepsilon_N\le K_P^+/N$.
If $K_P^+/N\le1/(8b^2)$, take $h=2\sqrt2\,b\sqrt{K_P^+/N}\le1$.
The two secant terms then sum to at most
$(2\sqrt2/b)\sqrt{K_P^+/N}$.
Otherwise $\delta_{M,N}\le\sum_iJ_iK_M(b_i)\le1/b$
and $1/b<2\sqrt2\sqrt{K_P^+/N}$.
Both cases give \eqref{prob:uniform-deficit}.
\end{proof}

\section{The upper bound}\label{main:upper}
\subsection{Full-vector domination, paths, and the Gram matrix}

In this subsection the field may be the above Gibbs field, or any field
on $\Lambda_M$ having translation invariance, simultaneous $O(N)$
invariance, the hard constraint $|S_x|^2=N$, and
\eqref{prob:GD}. The same statements hold on $\mathbb Z^d$ for
finitely supported zero-sum sources, with
$K_\infty(a)=\beta^{-1}Q_\infty(a)$ in place of $K_M(a)$,
$D_\infty(a,a)=K_\infty(a)-\mathbb E|\sum_xa_xS_x|^2/N$ in place of $D$,
and $\delta_{\infty,N}=\sum_iJ_iD_\infty(b_i,b_i)$ in place of
$\delta_{M,N}$.
The path argument uses this lattice geometry and translation invariance.

\begin{lemma}[Quantitative saturation of the full norm]
\label{prob:norm-saturation}
Under these assumptions $D$ is positive semidefinite on zero-sum
sources. Let $(\mathcal T_za)_x=a_{x-z}$ denote translation by
$z\in\mathbb Z^d$. For any path decomposition
$a=\sum_q\alpha_q\mathcal T_{z_q}b_{i_q}$,
\begin{equation}
 D(a,a)\le\delta_{M,N}
       \left(\sum_q\frac{|\alpha_q|}{\sqrt{J_{i_q}}}\right)^2.
 \label{prob:path-bound}
\end{equation}
For $Y_a=\sum_xa_xS_x$, $U_a=|Y_a|^2/N$ and every $N\ge1$,
\begin{equation}
 \mathbb E|U_a-K_M(a)|
 \le D(a,a)+c_\chi K_M(a)/\sqrt N,
 \qquad c_\chi=2\sqrt\pi+4.
 \label{prob:norm-L1}
\end{equation}
When $K_M(a)>0$, the sharper upper-tail bound is
\begin{equation}
 \mathbb P\{U_a\ge K_M(a)(1+s)\}
 \le\exp\{-N[s-\log(1+s)]/2\},\qquad s>0.
 \label{prob:chi-tail}
\end{equation}
\end{lemma}

\begin{proof}
Rotation invariance gives $\mathbb EY_a=0$ and
$\mathbb E(Y_a^1)^2=\mathbb E|Y_a|^2/N$.
Taking the second derivative at zero in \eqref{prob:GD} proves
$D(a,a)\ge0$.
Since $D$ is a difference of quadratic forms, its polarization satisfies
$|D(a,b)|\le\sqrt{D(a,a)D(b,b)}$.
Translation invariance and positivity of the $J_i$ give
$D(\mathcal T_zb_i,\mathcal T_zb_i)=D(b_i,b_i)\le\delta_{M,N}/J_i$.
The triangle inequality for this positive semidefinite seminorm gives
\eqref{prob:path-bound}.

Let $g\sim\mathcal N(0,I_N)$ be independent.  Tonelli's theorem and
\eqref{prob:GD} give
\begin{equation}
 \mathbb E e^{tU_a/2}
 =\mathbb E_g\mathbb E e^{\sqrt{t/N}\,g\cdot Y_a}
 \le(1-tK_M(a)/N)^{-N/2},
 \quad 0\le t<N/K_M(a).
 \label{prob:linearization}
\end{equation}
If $K_M(a)=0$, the second-moment bound makes $Y_a=0$ almost surely,
and \eqref{prob:norm-L1} is immediate.
Otherwise put $K=K_M(a)$.
Optimizing the Chernoff bound in \eqref{prob:linearization} at
$t=(N/K)s/(1+s)$ proves \eqref{prob:chi-tail}.
For an explicit integrated bound, the inequality
$-\log(1-z)-z\le z^2/[2(1-z)]$, $0\le z<1$, gives
\[
 \log\mathbb E e^{\theta(U_a-K)}
 \le\frac{v\theta^2}{2(1-c\theta)},
 \quad v=2K^2/N,\quad c=2K/N,\quad 0\le\theta<1/c.
\]
For $x>0$ use
$\theta=\sqrt{2x/v}/(1+c\sqrt{2x/v})$.
The resulting Chernoff bound is
\[
 \mathbb P\{U_a-K\ge2K\sqrt{x/N}+2Kx/N\}\le e^{-x}.
\]
Integration after this increasing change of threshold yields
\[
 \mathbb E(U_a-K)_+
 \le K\int_0^\infty e^{-x}
           \left(\frac1{\sqrt{Nx}}+\frac2N\right)dx
 =K\left(\frac{\sqrt\pi}{\sqrt N}+\frac2N\right).
\]
Finally
$\mathbb E(K-U_a)_+-\mathbb E(U_a-K)_+=D(a,a)$.
Adding the two positive parts and using $N^{-1}\le N^{-1/2}$ proves
\eqref{prob:norm-L1}.
\end{proof}

\begin{lemma}[Gram error and a positive target]
\label{prob:Gram}
Fix distinct sites $\mathcal U=\{x_1,\ldots,x_\ell\}$, injectively embedded
in the torus, and put
$A_{ij}=S_{x_i}\cdot S_{x_j}/N$ and
$a_{ij}=\delta_{x_i}-\delta_{x_j}$.
For each pair choose a fixed nearest-neighbour path and let
$h_{ij}(J)$ be the sum of $J_q^{-1/2}$ over its steps.
Suppose a deterministic matrix $\Sigma$ satisfies
\[
 \Sigma_{ii}=1,\qquad \Sigma\succeq mI_\ell\quad(m>0),\qquad
 \Sigma_{ij}=1-\tfrac12K_\infty(a_{ij}).
\]
Define
\[
 H_{\mathcal U,J}=\frac12\sum_{i,j}h_{ij}(J)^2,
 \quad R_{\mathcal U,M}=\frac{c_\chi}{2}\sum_{i,j}K_M(a_{ij}),
 \quad \epsilon_{\mathcal U}(M)=\frac12\sum_{i,j}
                      |K_M(a_{ij})-K_\infty(a_{ij})|.
\]
Then
\begin{equation}
 \mathbb E\|A-\Sigma\|_F
 \le H_{\mathcal U,J}\delta_{M,N}
           +R_{\mathcal U,M}N^{-1/2}+\epsilon_{\mathcal U}(M).
 \label{prob:Gram-error}
\end{equation}
For the critical or ordered lattice target
\[
 \Sigma_{ij}=\frac{G_J(x_i,x_j)}\beta+1-\frac{\beta_{c,J}}\beta,
 \qquad \beta\ge\beta_{c,J},
 \quad \beta_{c,J}=G_J(0,0),
\]
one has $K_\infty(a)=\beta^{-1}(a,G_Ja)$ and
\begin{equation}
 \Sigma\succeq(4\beta j)^{-1}I_\ell.
 \label{prob:target-positive}
\end{equation}
\end{lemma}

\begin{proof}
The full Gram matrix has $A\succeq0$, $A_{ii}=1$, and
$\operatorname{tr}A=\ell$.
Hard length gives $A_{ij}=1-U_{a_{ij}}/2$.
Apply \eqref{prob:path-bound} and \eqref{prob:norm-L1} entrywise,
then use $\|B\|_F\le\sum_{i,j}|B_{ij}|$ to prove
\eqref{prob:Gram-error}.
For the specified lattice target the dipole relation follows from
translation invariance of $G_J$.
If $z\in\mathbb R^\ell$ and $a=\sum_i z_i\delta_{x_i}$, Parseval and
$\lambda_J\le4j$ almost everywhere give
\[
 z^{\mathsf T}\Sigma z
 =\frac1\beta\int\frac{|\widehat a(k)|^2}{\lambda_J(k)}
                    \frac{dk}{(2\pi)^d}
  +\left(1-\frac{\beta_{c,J}}\beta\right)(\sum_i z_i)^2
 \ge\frac{|z|^2}{4\beta j}.
\]
The integrals exist for $d\ge3$.
For uniform constants on $J\in[j_0,j_1]^d$, $\beta\in[b,B]$, let
$n_{ij}$ be the chosen path's step count.
Then $h_{ij}\le n_{ij}/\sqrt{j_0}$, and Fourier path decomposition gives
$|\widehat a_{ij}(k)|\le h_{ij}\sqrt{\lambda_J(k)}$.
Consequently
\[
 H_{\mathcal U,J}\le H_*:=\frac1{2j_0}\sum_{i,j}n_{ij}^2,
 \qquad R_{\mathcal U,M}\le R_*:=c_\chi H_*/b.
\]
For the critical or ordered lattice target one may take
$m=m_*:=(4Bdj_1)^{-1}$ by \eqref{prob:target-positive}.
The remaining deterministic term is a fixed finite collection of
zero-sum Green errors, estimated in Lemma~\ref{spec:quadrature}.
\end{proof}

\subsection{Haar projection with all admissible coordinate counts}

\begin{proposition}[Quantitative rotational projection]
\label{prob:Haar}
Let $X\in\mathbb R^{\ell\times N}$ be invariant in law under right
multiplication by every $O\in O(N)$, and let $A=XX^{\mathsf T}/N$ satisfy
$A_{ii}=1$ almost surely.
If $N\ge\ell$, $1\le p\le N$, and $\Sigma\succeq mI_\ell$ is deterministic for
some $m>0$, then the law $\nu^{\mathcal U,p}$ of the first $p$ columns satisfies
\begin{equation}
 W_1(\nu^{\mathcal U,p},\gamma_\Sigma^{\otimes p})
 \le\sqrt p\left[
      \frac{\kappa_\ell}{\sqrt N}
       +m^{-1/2}\mathbb E\|A-\Sigma\|_F\right],
 \qquad \kappa_\ell=(\sqrt2+4)\ell^{3/2}\sqrt{\ell+1}.
 \label{prob:Haar-bound}
\end{equation}
Here $W_1$ uses the unnormalized Frobenius norm and
$\gamma_\Sigma^{\otimes p}$ has $p$ independent $\mathcal N(0,\Sigma)$ columns.
The random Gram matrix $A$ may be singular.
\end{proposition}

\begin{proof}
Let $P=[I_\ell\ 0]$ and let $E_p:\mathbb R^p\to\mathbb R^N$ embed the
first $p$ coordinates.
Every deterministic $X$ has the same row Gram matrix as
$\sqrt N A^{1/2}P$, hence belongs to its right orthogonal orbit.
For a rank-deficient $X$, the isometry between the two row spaces
extends to an orthogonal map on $\mathbb R^N$, so this assertion still
holds.
If $H$ is an independent Haar-distributed random matrix in $O(N)$,
rotational invariance
and Haar averaging therefore give the exact kernel representation
\[
 XE_p\ \overset{\mathrm{law}}=\ A^{1/2}Q_N,
 \qquad Q_N=\sqrt N\,PHE_p,
 \qquad Q_N\text{ independent of }A.
\]
This follows by averaging each deterministic orbit.

Independently of $A$, take $G\in\mathbb R^{N\times\ell}$ with independent
standard normal entries, and set
\[
 B=G^{\mathsf T}G/N,
 \quad Z=G^{\mathsf T}E_p,
 \quad T=G(G^{\mathsf T}G)^{-1/2}.
\]
Since $N\ge\ell$, $G$ has full column rank almost surely.
The law of $T$ is left $O(N)$ invariant on
$\{T:T^{\mathsf T}T=I_\ell\}$, so it is the Haar Stiefel law.
Thus we may realize the preceding kernel as
\[
 Q_N=\sqrt N\,T^{\mathsf T}E_p=B^{-1/2}Z.
\]
The entries of $Z$ remain independent standard normals for every
$1\le p\le N$; $B$ and $Z$ are in general dependent.
The entire coupling $(B,Z,Q_N)$ is independent of $A$.
Direct second-moment computations yield
\[
 \mathbb E\|B-I_\ell\|_F^2=\frac{\ell(\ell+1)}N,
 \qquad \mathbb E\|Z\|_F^2=\mathbb E\|Q_N\|_F^2=\ell p.
\]
On $\mathcal E=\{\|B-I_\ell\|_{\mathrm{op}}\le1/2\}$,
spectral calculus gives
$\|B^{-1/2}-I\|_{\mathrm{op}}
 \le\sqrt2\|B-I\|_{\mathrm{op}}$.
Cauchy--Schwarz then bounds the good-event expectation by
\[
 \mathbb E[\|Q_N-Z\|_F;\mathcal E]
 \le\sqrt2\sqrt{\ell(\ell+1)/N}\sqrt{\ell p}.
\]
Markov gives $\mathbb P(\mathcal E^c)\le4\ell(\ell+1)/N$.
On its complement we use only the second moments:
\[
 \mathbb E[\|Q_N-Z\|_F;\mathcal E^c]
 \le(\sqrt{\mathbb E\|Q_N\|_F^2}
            +\sqrt{\mathbb E\|Z\|_F^2})
                 \sqrt{\mathbb P(\mathcal E^c)}
 \le4\sqrt{\ell p\,\ell(\ell+1)/N}.
\]
Adding the bounds gives
\begin{equation}
 \mathbb E\|Q_N-Z\|_F
 \le(\sqrt2+4)\sqrt{\ell p\,\ell(\ell+1)/N}.
 \label{prob:Haar-good-bad}
\end{equation}

The square-root comparison uses only the lower bound on $\Sigma$:
\begin{equation}
 \|A^{1/2}-\Sigma^{1/2}\|_F\le m^{-1/2}\|A-\Sigma\|_F.
 \label{prob:sqrt-Lipschitz}
\end{equation}
Indeed, for $S=A^{1/2}-\Sigma^{1/2}$,
$A-\Sigma=A^{1/2}S+S\Sigma^{1/2}$.
In the left eigenbasis of $A^{1/2}$ and the right eigenbasis of
$\Sigma^{1/2}$ the entries are related by the factors
$\sqrt{a_i}+\sqrt{c_j}\ge\sqrt m$.
The Frobenius norm is invariant under these two changes of basis,
proving \eqref{prob:sqrt-Lipschitz} even for singular $A$.
Finally $\|A^{1/2}\|_{\mathrm{op}}\le\sqrt\ell$, and independence of
$A$ and $Z$ implies
\[
 \mathbb E_Z\|(A^{1/2}-\Sigma^{1/2})Z\|_F
 \le\sqrt p\,\|A^{1/2}-\Sigma^{1/2}\|_F.
\]
Couple $A^{1/2}Q_N$ to $\Sigma^{1/2}Z$ and combine the last inequalities
with \eqref{prob:Haar-good-bad} to obtain \eqref{prob:Haar-bound}.
\end{proof}

\begin{corollary}[Quantitative saturation criterion]
\label{prob:saturation-criterion}
Under the assumptions of Lemma~\ref{prob:Gram}, for $N\ge\ell$ and
$1\le p\le N$,
\begin{equation}
 W_1(\nu^{\mathcal U,p},\gamma_\Sigma^{\otimes p})
 \le\sqrt p\left[
 \frac{\kappa_\ell+m^{-1/2}R_{\mathcal U,M}}{\sqrt N}
 +m^{-1/2}H_{\mathcal U,J}\delta_{M,N}
 +m^{-1/2}\epsilon_{\mathcal U}(M)\right].
 \label{prob:saturation-bound}
\end{equation}
For an infinite-lattice field satisfying the same assumptions with
$K_\infty$, the last term is absent, $\delta_{M,N}$ is replaced by
$\delta_{\infty,N}$, and $R_{\mathcal U,M}$ is replaced by
$R_{\mathcal U,\infty}=(c_\chi/2)\sum_{i,j}K_\infty(a_{ij})$.
All displayed dependence on $p$ is the factor $\sqrt p$.
\end{corollary}

\begin{proof}
Insert \eqref{prob:Gram-error} into \eqref{prob:Haar-bound}.
For the infinite field the proof of the Gram bound uses the reference
form directly, so the deterministic reference error is zero.
For the critical or ordered lattice target, the bounds in
Lemma~\ref{prob:Gram} make these constants uniform over
\eqref{main:compact}.
\end{proof}

\subsection{Proof of the uniform approximation}
\begin{proof}[Proof of Theorem~\ref{main:joint}]
For the full-spin Gram matrix
\(A_{uv}=S_{x_u}\cdot S_{x_v}/N\), hard length gives \(A_{uu}=1\) and
\[
 A_{uv}=1-\frac1{2N}|S_{x_u}-S_{x_v}|^2.
\]
In the torus Gibbs model, combining \eqref{prob:uniform-deficit} with
the sharp zero-sum quadrature estimate
\(Q_M^0(a)-Q_\infty(a)=O(M^{-d})\) in
Lemma~\ref{spec:quadrature} gives
\[
 \E\|A-\Sigma_{\beta,J}\|_F
 \le C_\U(N^{-1/2}+s_M+M^{-d})
\]
by Lemma~\ref{prob:Gram}.
By \eqref{prob:target-positive},
\(\Sigma_{\beta,J}\succeq(4\beta j)^{-1}I_\ell\).
Proposition~\ref{prob:Haar}, valid even when \(A\) is singular, therefore
gives (this is Corollary~\ref{prob:saturation-criterion})
\[
 W_1\le C_{\mathcal U}\sqrt p
           (N^{-1/2}+s_M+M^{-d}).
\]
The mass equation gives \(s_M\ge1/(\beta V)\), so \(M^{-d}\le Bs_M\) on
\(\beta\le B\); this proves \eqref{main:bound}.
\end{proof}

\section{Lower bounds and the coordinate criterion}\label{main:lower}
The lower bounds below concern the actual finite-spin laws. The
argument uses radial symmetry and both the infrared and pressure
comparisons.

\subsection{The coordinate-growth obstruction}
The coordinate-growth obstruction below comes from the one-site
sphere marginal. Related sharp one-dimensional Gaussian asymptotics
for Euclidean-ball marginals are studied in
\cite{ChenHuang}.

\begin{proposition}[Necessary coordinate-growth condition]
\label{prob:coordinate-obstruction}
Let $N\to\infty$, $1\le p_N\le N$, and let the site set be fixed and
nonempty.  For any globally $O(N)$ invariant spin laws with $|S_x|^2=N$,
if $p_N/N\to c\in(0,1]$, then comparison to a centred Gaussian block with
independent columns and unit diagonal covariance satisfies
\begin{equation}
 \liminf_N W_1(\nu_N^{\mathcal U,p_N},\gamma_N^{\mathcal U,p_N})
 \ge\frac{1-\sqrt{1-c}}{\sqrt\pi}>0.
 \label{prob:coordinate-necessary}
\end{equation}
Thus $p_N=o(N)$ is necessary for convergence in the unnormalized
Euclidean $W_1$ metric.
\end{proposition}

\begin{proof}
It suffices to project onto one site, a $1$-Lipschitz map.
Let $\Theta$ be the norm of its first $p=p_N$ spin coordinates, and $T$ the
norm of a standard $p$-dimensional Gaussian vector.
The Gaussian representation of uniform directions gives
\[
 \Theta^2\ \overset{\mathrm{law}}=\frac{N A_p}{A_p+B_{N-p}},
 \qquad A_p\sim\chi_p^2,\quad B_{N-p}\sim\chi_{N-p}^2
 \quad\hbox{independently},
\]
where $B_0=0$.
Thus $\mathbb E\Theta^2=p$ and
$\operatorname{Var}(\Theta^2)=2p(N-p)/(N+2)$, whence
\[
 \mathbb E(\Theta-\sqrt p)^2
 \le\mathbb E(\Theta^2-p)^2/p=2(N-p)/(N+2)\le2.
\]
If $0<c<1$, the two chi-square central limit theorems and expansion of
$\sqrt{N A_p/(A_p+B_{N-p})}$ at $(p,N-p)$ give
\[
 \Theta-\sqrt p
 =\frac{1-p/N}{2\sqrt p}(A_p-p)
  -\frac{p/N}{2\sqrt p}(B_{N-p}-(N-p))+o_{\mathbb P}(1)
 \ \Longrightarrow\ \mathcal N(0,(1-c)/2).
\]
Likewise $T-\sqrt p\Longrightarrow\mathcal N(0,1/2)$ and
$\mathbb E(T-\sqrt p)^2\le\operatorname{Var}(T^2)/p=2$.
These second-moment bounds make the absolute centred norms uniformly
integrable, so their expectations tend respectively to
$\sqrt{1-c}/\sqrt\pi$ and $1/\sqrt\pi$.
If $c=1$, the bound for $\Theta-\sqrt p$ tends to zero in $L^2$, giving the
same formula.
The function $u\mapsto\bigl||u|-\sqrt p\bigr|$ is $1$-Lipschitz.
Its expectation difference proves \eqref{prob:coordinate-necessary}.
If $p_N/N$ does not tend to zero, a subsequence has a positive limiting
ratio, which rules out convergence.
\end{proof}

\subsection{Star lower bounds and the coordinate criterion}
\begin{proposition}[Finite-spin star lower bounds]\label{floor:bound}
Under \eqref{main:compact}, for every integer $M\ge4$, $N\ge1$
and $1\le p\le N$, with $\mathcal U_0=\{0,e_1,\ldots,e_d\}$, write
\[
 w=W_1(\nu_{M,N,\beta,J}^{\mathcal U_0,p},
               \gamma_{\beta,J}^{\mathcal U_0,p}).
\]
Then
\begin{equation}\label{floor:lower}
 w\ge\frac{\sqrt p}{4\sqrt\pi(1+\sqrt2)\beta jV}.
\end{equation}
For $N\ge4$, let
\[
 K_P^-=8(1+4Bdj_1)+\frac32,\qquad C_E=(8K_P^-+1)/b.
\]
We also have
\begin{equation}\label{floor:mass-lower}
 w\ge\frac{\sqrt p}{4\sqrt\pi(1+\sqrt2)j}
       \max\left\{\frac1{\beta V},\ s_M-\frac{C_E}{\sqrt N}\right\}.
\end{equation}
In particular $w\ge c\sqrt p\,s_M-C\sqrt{p/N}$ uniformly on the
compact parameter ranges. Both lower bounds hold for any fixed window
containing a translated star. If $N^{-1/2}\le s_M/(2C_E)$, the mass
order $\sqrt p\,s_M$ is matched at finite spin dimension.
\end{proposition}
\begin{proof}
Let $v_i=\mathbb E(S_0^1-S_{e_i}^1)^2$ and let $v_i^\infty$ be the
corresponding single-column target variance. Hard length and the unit
target diagonal give $0\le v_i,v_i^\infty\le4$. Set
\[
 \omega=\sum_iJ_i(v_i^\infty-v_i).
\]
The target constant mode cancels from each dipole. The exact Fourier
identities and the spin energy derivative give
\[
 \sum_iJ_iv_i^\infty=\frac1\beta,\qquad
 \sum_iJ_iv_i=2(j-P_{M,N}'),\qquad
 \omega=s_M+2[P_{M,N}'-(P_M^{\mathrm{comp}})'].
\]
The last equality uses Lemma~\ref{prob:energy-deficit}.
Gaussian domination bounds
$v_i\le Q_M^0(\delta_0-\delta_{e_i})/\beta$; hence
\[
 \omega\ge\frac1{\beta V}>0,
\]
because $\sum_iJ_iQ_M^0(\delta_0-\delta_{e_i})=1-V^{-1}$; by
\eqref{prob:energy-identity}, \(\omega=\delta_{M,N}+1/(\beta V)\).

We first turn $\omega$ into a Wasserstein lower bound. Write
$\bar\chi_q=\mathbb E\chi_q$ and $d_N=\sqrt N/\bar\chi_N\ge1$, where $\chi_q$ is the
norm of a standard Gaussian vector in $\mathbb R^q$. For any rotationally
invariant $Y\in\mathbb R^N$, Haar averaging gives the exact identity
\[
 \mathbb E|Y^{[p]}|=\frac{\bar\chi_p}{\bar\chi_N}\mathbb E|Y|.
\]
Indeed, a standard Gaussian radius is independent of its uniform
direction, so the expected norm of the first $p$ coordinates of a
uniform unit vector is $\bar\chi_p/\bar\chi_N$. Averaging a fixed vector of any radius
proves the identity, including a possible atom at zero and $N=1$.
The one-site spin marginal is exactly uniform on the sphere of radius
$\sqrt N$. Thus
\[
 \mathbb E|S_0^{[p]}|=d_N\bar\chi_p,\qquad
 \mathbb E|(S_0-S_{e_i})^{[p]}|
 \le d_N\bar\chi_p\sqrt{v_i},
\]
where the inequality follows from Jensen and
$\mathbb E|S_0-S_{e_i}|^2=Nv_i$.

Choose a direction with $v_i^\infty-v_i\ge \omega/j>0$. On the star block
$z=(z_0,z_{e_1},\ldots,z_{e_d})$, consider
\[
 F_i(z)=\sqrt{v_i^\infty}\,|z_0|-|z_0-z_{e_i}|.
\]
Its Lipschitz constant for the Frobenius norm is at most $2+\sqrt2$.
Its expectation under the target is zero: the two expected norms are
$\bar\chi_p$ and $\bar\chi_p\sqrt{v_i^\infty}$, respectively. Under the spin law,
\[
 \mathbb EF_i
 \ge d_N\bar\chi_p(\sqrt{v_i^\infty}-\sqrt{v_i})
 \ge\frac{\bar\chi_p(v_i^\infty-v_i)}4.
\]
Here $\sqrt{v_i^\infty}+\sqrt{v_i}\le4$. The elementary inequality
$|g|\ge p^{-1/2}\sum_{\alpha=1}^p|g_\alpha|$ gives
$\bar\chi_p\ge\sqrt{2p/\pi}$. Kantorovich duality therefore yields
\[
 w\ge\frac{\sqrt p\,\omega}{4\sqrt\pi(1+\sqrt2)j}.
\]
Combining this with $\omega\ge1/(\beta V)$ proves \eqref{floor:lower} for
every finite spin dimension.

For the mass lower bound take $N\ge4$ and $h=b/(2\sqrt N)$.
The interval $[\beta-h,\beta]$ lies in $[b/2,B]$. Proposition
\ref{prob:pressure} applies on this interval even when its left end is
below $\beta_{c,J}$, and gives a pressure error at most $K_P^-/N$.
Convexity of $P_{M,N}$ and the curvature bound from
Lemma~\ref{prob:energy-deficit} imply
\[
 \begin{split}
 P_{M,N}'(\beta)
 &\ge\frac{P_{M,N}(\beta)-P_{M,N}(\beta-h)}h\\
 &\ge(P_M^{\mathrm{comp}})'(\beta)
       -\frac{2K_P^-}{Nh}-\frac h{b^2}.
 \end{split}
\]
The last term follows by integrating
$(P_M^{\mathrm{comp}})''\le2/b^2$ to bound the difference between
the derivative at $\beta$ and the secant on $[\beta-h,\beta]$. Substitution of $h$ gives
\[
 \omega=s_M+2[P_{M,N}'-(P_M^{\mathrm{comp}})']
   \ge s_M-\frac{8K_P^-+1}{b\sqrt N}.
\]
Together with $\omega\ge1/(\beta V)$ and the norm-test bound, this proves
\eqref{floor:mass-lower}. Projection onto a contained star is $1$-Lipschitz;
translation invariance supplies the claim for translated stars.
\end{proof}

\begin{proof}[Proof of Theorem~\ref{main:finite-criterion}]
The two finite-spin lower estimates are \eqref{floor:lower}, since
\(\beta\le B\) and \(j\le dj_1\) on \eqref{main:compact}, and the
consequence \(w\ge c\sqrt p\,s_M-C\sqrt{p/N}\) of
\eqref{floor:mass-lower} stated in Proposition~\ref{floor:bound}.
If the joint distance tends to zero, projection onto one site and the
sphere obstruction of Proposition~\ref{prob:coordinate-obstruction} force
$p_n=o(N_n)$. The mass lower bound then gives
\[
 c\sqrt{p_n}\,s_{M_n,\beta_n,J_n}
 \le W_1+C\sqrt{p_n/N_n}\longrightarrow0.
\]
Conversely, the two conditions make the upper bound in
Theorem~\ref{main:joint} tend to zero. A projection onto a translated
star is $1$-Lipschitz, so the same necessity holds on every fixed
window containing such a star. By the uniform mass comparison in
Proposition~\ref{spec:crossover}, the volume condition is equivalent to
\[
 p_n=o\!\left([V_n\{\beta_n-\beta_{c,J_n}+t_d(M_n)\}]^2\right).
\]
This gives the displayed conditions at criticality and under a
uniform positive ordered gap.
\end{proof}

\subsection{Periodic accumulation states}
\begin{proof}[Proof of Corollary~\ref{main:infinite}]
At fixed \(N\), extend each torus configuration periodically to
\(\Z^d\). The product of the compact spin spheres on this countable
site set is a compact metrizable space, so these laws have local weak
accumulation points. Translation and global rotation invariance pass
to the limit. These accumulation states satisfy the
Dobrushin--Lanford--Ruelle
equations of the infinite-lattice Gibbs specification. For a fixed
finite region, test the finite-torus conditional identity with a
continuous function of the spins in that region and a continuous
exterior cylinder function. For all sufficiently large \(M\), the
region, its interaction boundary and the test supports embed without
identifications in the torus. The conditional kernel then agrees with
that on \(\Z^d\) and is continuous in the boundary spins. The identity
passes to the weak limit; cylinder approximation and the monotone-class
theorem extend it to bounded measurable tests.
Each coordinate has second moment one, so the local first moments are
uniformly integrable and the finite-volume bound passes to \(W_1\).
For \(N\ge\max(4,\ell)\), send \(M\to\infty\) in
Theorem~\ref{main:joint}. For \(N<\max(4,\ell)\) and a unit-diagonal
target, the independent coupling bound is \(W_1\le2\sqrt{\ell p}\);
enlarging the uniform constant by \(2\sqrt{\ell\max(4,\ell)}\) extends
the \(C_{\mathcal U}\sqrt{p/N}\) statement to all \(N\ge1\), \(p\le N\).
If \(p_N/N\) fails to tend to zero, it has a subsequence along
which \(p_N/N\to c\in(0,1]\). On that subsequence
Proposition~\ref{prob:coordinate-obstruction} gives
\(\liminf W_1\ge(1-\sqrt{1-c})/\sqrt\pi>0\).
\end{proof}

\section{The ordered regime}\label{main:ordered}
Throughout this section the parameters lie in the compact ranges
\eqref{main:compact}. We additionally fix $\kappa>0$ and assume
$\beta-\beta_{c,J}\ge\kappa$. Put $L_N=\log(\mathrm e N)$, and let
$\psi_d(N)$ be as in Theorem~\ref{main:ordered-rate}.
Constants are uniform in $M,N,\beta,J$; constants in the local
estimates may also depend on the fixed window.
Sections~\ref{ordered:sec-fluct}--\ref{ordered:sec-ward} prove the
deficit bound of Proposition~\ref{ordered:deficit};
Section~\ref{ordered:sec-mixture} gives the mixture lemma and the
estimate in terms of \(\delta_{M,N}\); Section~\ref{ordered:sec-large}
proves Proposition~\ref{main:sector}; and
Section~\ref{ordered:sec-proof} proves
Theorem~\ref{main:ordered-rate}.

\subsection{One-sided quadratic fluctuations}\label{ordered:sec-fluct}
For a zero-sum source $a$ on the torus, write
\[
 U_a=\frac1N\left|\sum_xa_xS_x\right|^2,
 \qquad K_a=\beta^{-1}Q_M^0(a),\qquad D_a=K_a-\E U_a.
\]
The source may have support depending on $M$.
For a real variable $F$, let $F^+=\max(F,0)$ and
$F^-=\max(-F,0)$.
The event estimate below keeps the mean deficit $D_a$ linear while
allowing the event to depend on a second source. Through the layer-cake
argument it controls the products $F_{b_{xy}}F_{a_z}$ used in the Ward
identity argument of Section~\ref{ordered:sec-ward}.
\begin{lemma}[Quadratic fluctuations and their upper tail]
\label{ordered:one-sided}
For $N\ge2$ and $F_a=K_a-U_a$,
\begin{align}
 \E F_a^2&\le4\mathrm e^{1/2}K_aD_a+8\mathrm e^{1/2}K_a^2/N,
 \label{ordered:quadratic}\\
 \E(F_a^-)^2&\le C K_a^2/N.
 \label{ordered:negative}
\end{align}
In particular $\E(F_a^+)^2\le C(K_aD_a+K_a^2/N)$.
For every event $E$,
\begin{equation}
 \E[F_a^+\mathbf1_E]\le \mathrm e\,D_a+CK_a\sqrt{\mathbb P(E)/N}.
 \label{ordered:event}
\end{equation}
If $K_a,K_c\le K_*$ for two zero-sum sources, then
\begin{equation}
 -\E(F_aF_c)\le C_{K_*}\left\{
   \sqrt{L_N/N}(D_a+D_c)+N^{-1}\right\}.
 \label{ordered:mixed}
\end{equation}
\end{lemma}
\begin{proof}
If $K_a=0$, domination forces $U_a=0$ almost surely. Otherwise Gaussian
linearization of \eqref{prob:GD} gives
\[
 \E e^{tU_a/(2K_a)}\le(1-t/N)^{-N/2},\qquad 0\le t<N.
\]
Set $X=(U_a-K_a)/(2K_a)\ge-1/2$. At $t=1$ the bound is
$\E e^X\le e^{1/[4(N-1)]}\le1+1/N$.
The integral remainder in the exponential gives
$e^x-1-x\ge e^{-1/2}x^2/2$ for $x\ge-1/2$.
Using $\E X=-D_a/(2K_a)$ proves \eqref{ordered:quadratic}.
For \eqref{ordered:negative}, take $t=\sqrt N/2$. Then
\[
 \log\E e^{tX}
 \le-\frac t2-\frac N2\log(1-t/N)
 \le\frac{t^2}{4N(1-t/N)}\le\frac14.
\]
Consequently
$\mathbb P(U_a-K_a>u)\le e^{1/4}e^{-\sqrt N u/(4K_a)}$.
Integrating $2u$ times this tail over $u>0$ proves
\eqref{ordered:negative}. This uses domination for a single source only.

For \eqref{ordered:event}, first suppose $N\ge16$ and $K_a>0$.
Put $f=F_a/K_a$ and $d_a=D_a/K_a$. For $0\le\lambda\le\sqrt N$,
Gaussian linearization and the logarithmic remainder bound give
\[
 \E e^{-\lambda f}\le e^{H_N(\lambda)},\qquad
 H_N(\lambda)=-\lambda-\frac N2\log(1-2\lambda/N)
 \le\frac{\lambda^2}{N(1-2\lambda/N)}
 \le\frac{2\lambda^2}N\le2.
\]
The convex remainder $e^{-\lambda f}-1+\lambda f$ is nonnegative
for every real $f$, and hence
\[
 \E\{e^{-\lambda f}-1+\lambda f\}
 \le(e^2-1)\lambda^2/N+\lambda d_a.
\]
For $y\ge1$, $e^{-y}-1+y\ge e^{-1}y$; indeed the ratio of the
left side to $y$ is increasing there and equals $e^{-1}$ at one.
For $s\ge N^{-1/2}$ take $\lambda=1/s$ and restrict the nonnegative
remainder to $\{f>s\}$. This proves
\[
 \E[f^+\mathbf1_{\{f>s\}}]\le ed_a+C/(Ns).
\]
If $q=\mathbb P(E)>0$, choose $s=(Nq)^{-1/2}$ and split according
to $f\le s$ and $f>s$. Then
\[
 \E[f^+\mathbf1_E]\le sq+ed_a+C/(Ns)
                  \le ed_a+C\sqrt{q/N}.
\]
For $q=0$ the claim is immediate. For $2\le N\le15$, use
$0\le f^+\le1$ and $q\le\sqrt{15}\sqrt{q/N}$.
The case $K_a=0$ was already treated by domination. Multiplication
by $K_a$ proves \eqref{ordered:event}.

For \eqref{ordered:mixed}, the Chernoff tail bound in the proof of
Lemma~\ref{prob:norm-saturation} gives, with
\[
 T_a=2K_a\sqrt{3L_N/N}+6K_aL_N/N\le C K_a\sqrt{L_N/N},
\]
$\mathbb P(F_a^->T_a)\le e^{-3L_N}$.
By \eqref{ordered:negative} and Cauchy--Schwarz,
$\E[F_a^-\mathbf1_{\{F_a^->T_a\}}]\le C K_a N^{-2}$.
Since $0\le F_c^+\le K_c$ and
the upper-tail bound used for \eqref{ordered:negative} gives
$q_a(u):=\mathbb P(F_a^->u)\le
e^{1/4}e^{-\sqrt N u/(4K_a)}$, we have
$\int_0^\infty\sqrt{q_a(u)}\,du\le CK_a/\sqrt N$.
Use the layer-cake identity and \eqref{ordered:event} on the possibly
correlated event $\{F_a^->u\}$:
\[
\begin{aligned}
 \E(F_a^-F_c^+)
 &\le\int_0^{T_a}\E[F_c^+\mathbf1_{\{F_a^->u\}}]\,du
       +K_c\E[F_a^-\mathbf1_{\{F_a^->T_a\}}]\\
 &\le \mathrm e\,D_cT_a+\frac{CK_c}{\sqrt N}
              \int_0^{T_a}\sqrt{q_a(u)}\,du+CK_aK_cN^{-2}\\
 &\le C K_aD_c\sqrt{L_N/N}+CK_aK_c/N.
\end{aligned}
\]
Interchange $a,c$ and discard the two nonnegative products in
$F_aF_c=F_a^+F_c^++F_a^-F_c^--F_a^+F_c^--F_a^-F_c^+$.
This proves \eqref{ordered:mixed}, including the cases of zero
reference forms by continuity or domination.
\end{proof}

\subsection{Magnetization and spectral interpolation}\label{ordered:sec-mag}
Define
\[
 \overline S=V^{-1}\sum_xS_x,\quad R=|\overline S|^2/N,
 \quad r=\E R,\quad t_x=S_x\cdot\overline S/N,
\]
\[
 a_x=\delta_x-V^{-1}\mathbf1,\quad
 K_0=\frac1{\beta V}\sum_{k\ne0}\lambda_J(k)^{-1},
 \quad r_0=1-K_0,\quad \eta=r-r_0.
\]
Here $K_{a_x}=K_0$, and hard length implies
\[
 U_{a_x}=1+R-2t_x,\qquad
 V^{-1}\sum_xU_{a_x}=1-R.
\]
Translation invariance gives $\E t_x=r$ and $D_{a_x}=\eta\ge0$.
The diagonal quadrature estimate in Lemma~\ref{spec:quadrature} yields
$K_0=\beta_{c,J}/\beta+O(M^{2-d})$.
Thus, after increasing a threshold depending on $\kappa$,
\begin{equation}
 r_0\ge\kappa/(2B)>0.
 \label{ordered:gap}
\end{equation}

Let $c(z)=\E(S_0\cdot S_z)/N$ and use the Fourier convention
$\widehat c(k)=\sum_{z\in\Lambda_M}c(z)e^{-ik\cdot z}$, so that
$V^{-1}\sum_k\widehat c(k)=c(0)=1$.
These Fourier eigenvalues are nonnegative. Gaussian domination gives,
for $k\ne0$,
\[
 d(k)=\frac1{\beta\lambda_J(k)}-\widehat c(k)\ge0,
 \qquad d(k)\le\frac1{\beta\lambda_J(k)}.
\]
With $\delta=\delta_{M,N}$ from Lemma~\ref{prob:energy-deficit},
\begin{equation}
 \eta=\frac1V\sum_{k\ne0}d(k),\qquad
 \delta=\frac1V\sum_{k\ne0}\lambda_J(k)d(k).
 \label{ordered:spectral-deficits}
\end{equation}
For centred representatives $k\in[-\pi,\pi]^d$,
$\lambda_J(k)\asymp |k|^2$ uniformly by the bounds on $J$.
For $0<u\le1$, counting shells $|k|\asymp n/M$ therefore gives
\[
 \frac1V\sum_{0<\lambda_J(k)\le u}\lambda_J(k)^{-1}
 \le C u^{(d-2)/2}.
\]
Indeed the shell sum is at most
$CM^{2-d}\sum_{n\le CM\sqrt u}n^{d-3}$.
If the low-frequency set is empty the estimate is immediate; otherwise
$M\sqrt u$ is bounded below and the displayed power follows.
Splitting \eqref{ordered:spectral-deficits} at $u$ gives
$\eta\le C u^{(d-2)/2}+\delta/u$.
For $0<\delta\le1$, take $u=\delta^{2/d}$. For $\delta=0$,
\eqref{ordered:spectral-deficits} gives $\eta=0$; larger $\delta$ are
covered by $\eta\le K_0\le C$. Hence
\begin{equation}
 \eta\le C\delta^{(d-2)/d}.
 \label{ordered:interpolation}
\end{equation}

\subsection{A Ward identity and the deficit bound}\label{ordered:sec-ward}
Let $J_{xy}=J_i$ when $y=x\pm e_i$, and put
\[
 h_x=\sum_{y\sim x}J_{xy}S_y,\qquad
 e_x=\sum_{y\sim x}J_{xy}(1-S_x\cdot S_y/N),
 \quad k_0=(1-V^{-1})/\beta.
\]
In particular $S_x\cdot h_x/N=2j-e_x$ and $\E e_x=k_0-\delta$.
On the sphere at $x$, let $P_x=I-S_xS_x^{\mathsf T}/N$.
The tangent vector field $P_x\overline S$ has divergence
$(N-1)/V-(N-1)t_x$. Integration by parts under the Gibbs density gives
\[
 0=(N-1)(V^{-1}-r)
   +\beta\E\{\overline S\cdot h_x-t_xS_x\cdot h_x\}.
\]
Translation invariance and $\sum_xh_x=2j\sum_xS_x$ turn the last
expectation into $N\E(t_xe_x)$. Therefore
\begin{equation}
 \beta r\delta=\beta\operatorname{Cov}(e_x,t_x)
                 +\frac rN+\frac{1-r-1/N}{V}.
 \label{ordered:ward}
\end{equation}

Set $F_x=K_0-U_{a_x}$, $\overline F=V^{-1}\sum_zF_z=R-r_0$,
and $f_x=k_0-e_x$. Then
\[
 t_x=r_0+\tfrac12(F_x+\overline F),\qquad
 \E f_x=\delta,\quad \E F_x=\E\overline F=\eta,
\]
so that
\begin{equation}
 \operatorname{Cov}(e_x,t_x)
 =\delta\eta-\tfrac12\E\{f_x(F_x+\overline F)\}.
 \label{ordered:cancellation}
\end{equation}
For $b_{xy}=\delta_x-\delta_y$,
$f_x=\frac12\sum_{y\sim x}J_{xy}F_{b_{xy}}$.
The reference equality
$\frac12\sum_{y\sim x}J_{xy}K_{b_{xy}}=k_0$ follows from
the Fourier trace of $L_{J,M}G^0_{J,M}$.
Likewise $\frac12\sum_{y\sim x}J_{xy}D_{b_{xy}}=\delta$.
All source reference forms are bounded uniformly.
Apply \eqref{ordered:mixed} separately to each pair
$b_{xy},a_z$, and then sum with the bond weights and, when needed,
the weights $1/V$. Since $D_{a_z}=\eta$, this gives, for
$G=F_x$ or $G=\overline F$,
\[
 \E(f_xG)\ge-C\{\sqrt{L_N/N}(\delta+\eta)+N^{-1}\}.
\]
Inserting this into \eqref{ordered:cancellation} and then
\eqref{ordered:ward} cancels $\delta\eta$ against the same part of
$r\delta=(r_0+\eta)\delta$:
\begin{equation}
 r_0\delta\le C\{\sqrt{L_N/N}(\delta+\eta)
                         +N^{-1}+V^{-1}\}.
 \label{ordered:closure}
\end{equation}
Here $0\le r\le1$ and $\beta\ge b$.

\begin{proposition}[Pointwise ordered deficit]
\label{ordered:deficit}
For $N\ge4$ and sufficiently large $M$, under
$\beta-\beta_{c,J}\ge\kappa$,
\begin{equation}
 \delta_{M,N}\le C_\kappa\{(L_N/N)^{d/4}
                        +N^{-1}+M^{-d}\}
 \le C_\kappa\{\psi_d(N)+M^{-d}\}.
 \label{ordered:deficit-rate}
\end{equation}
\end{proposition}
\begin{proof}
For sufficiently large $N$, absorb the
$C\sqrt{L_N/N}\,\delta$ term using \eqref{ordered:gap}.
Combine \eqref{ordered:interpolation} and \eqref{ordered:closure}.
Young's inequality, with any sufficiently small fixed
$\varepsilon>0$, gives
\[
 C\sqrt{L_N/N}\,\delta^{(d-2)/d}
 \le\varepsilon\delta+C_\varepsilon(L_N/N)^{d/4}.
\]
Absorb the linear term. The finitely many smaller $N\ge4$ are
covered by $0\le\delta\le1/b$. For $d\ge5$, the ratio
$N(L_N/N)^{d/4}=N^{1-d/4}L_N^{d/4}$ is bounded on $N\ge4$;
the cases $d=3,4$ follow directly. This proves
\eqref{ordered:deficit-rate}.
\end{proof}

\subsection{Gaussian mixtures and an estimate in terms of the deficit}\label{ordered:sec-mixture}
\begin{lemma}[Gaussian mixtures with explicit coordinate dependence]
\label{sector:mixture}
Fix $\ell$ and $m>0$. Let $A$ be a random positive semidefinite
$\ell\times\ell$ matrix of trace $\ell$, and let $\bar A=\E A\succeq mI$.
For an independent standard Gaussian matrix $Z\in\mathbb R^{\ell\times p}$,
\[
 W_1(\mathcal L(A^{1/2}Z),\mathcal L(\bar A^{1/2}Z))
 \le C_{\ell,m}p\,\E\|A-\bar A\|_F^2,
 \qquad p\ge1.
\]
If $\Sigma\succeq mI$ also has trace $\ell$, then
\[
 W_1(\mathcal L(\bar A^{1/2}Z),\mathcal L(\Sigma^{1/2}Z))
 \le C_m\sqrt p\,\|\bar A-\Sigma\|_F.
\]
\end{lemma}
\begin{proof}
For a $1$-Lipschitz function $g$ with $g(0)=0$, set
$\mathcal F_g(\Xi)=\E g(\Xi^{1/2}Z)$.
First suppose $g$ is smooth. On $mI/2\preceq \Xi\preceq\ell I$,
write $X=\Xi^{1/2}Z$ and, for a symmetric direction $H$, let
\[
 L_H(X)=-\frac p2\tr(\Xi^{-1}H)
       +\frac12\sum_{\alpha=1}^p
          X_\alpha^{\mathsf T}\Xi^{-1}H\Xi^{-1}X_\alpha.
\]
Differentiating the Gaussian density and integrating by parts once gives
\[
 D\mathcal F_g(\Xi)[H]=\frac12\E\sum_\alpha
       \nabla_\alpha g(X)\cdot H\Xi^{-1}X_\alpha.
\]
Since $\|\nabla g\|_F\le1$, this is bounded by
$C_m\sqrt p\,\|H\|_F$. Differentiating this expression with $X$
as the integration variable gives
\begin{align*}
 D^2\mathcal F_g(\Xi)[H,H]=\frac12\E\sum_\alpha\nabla_\alpha g(X)\cdot
 \{H\Xi^{-1}X_\alpha L_H(X)-H\Xi^{-1}H\Xi^{-1}X_\alpha\}.
\end{align*}
Here $\E L_H=0$ and
$\E L_H^2=(p/2)\tr(\Xi^{-1}H\Xi^{-1}H)$.
Also $\E\|H\Xi^{-1}X\|_F^2=p\tr(H\Xi^{-1}H)$.
Cauchy--Schwarz bounds the first term by $C_mp\|H\|_F^2$,
and the second by $C_m\sqrt p\,\|H\|_F^2$.
Thus $|D^2\mathcal F_g(\Xi)[H,H]|\le C_mp\|H\|_F^2$.
Smooth mollification and Gaussian moment bounds extend these formulas
and estimates to every Lipschitz $g$.

Put $H=A-\bar A$. If $\|H\|_{\rm op}\le m/2$, Taylor's theorem along
$\bar A+tH$ gives a remainder of at most $C_mp\|H\|_F^2$.
Otherwise $\|H\|_F>m/2$, while
$|\mathcal F_g(A)|,|\mathcal F_g(\bar A)|\le\sqrt{\ell p}$ and
$|D\mathcal F_g(\bar A)[H]|\le C_m\sqrt p\|H\|_F$.
These bounds give the same quadratic remainder with a constant
depending only on $\ell,m$. Taking expectations eliminates the
linear term because $\E H=0$. Kantorovich duality proves the first claim.
For the second, couple the two Gaussian matrices with the same $Z$.
The square-root estimate from Proposition~\ref{prob:Haar} gives
$\|\bar A^{1/2}-\Sigma^{1/2}\|_F\le C_m\|\bar A-\Sigma\|_F$; the Gaussian coupling
therefore has expected distance at most $C_m\sqrt p\|\bar A-\Sigma\|_F$.
\end{proof}

\begin{proposition}[Local comparison with explicit coordinate dependence]
\label{ordered:projection}
Fix $\mathcal U$. Under the assumptions of this section, for
sufficiently large $M$, $N\ge\max(4,\ell)$ and $1\le p\le N$, the
distance
$W_1=W_1(\nu_{M,N,\beta,J}^{\mathcal U,p},\gamma_{\beta,J}^{\mathcal U,p})$
satisfies
\begin{equation}\label{main:ordered-modular}
 W_1\le C_{\mathcal U}
       \left\{p(\delta_{M,N}+N^{-1})+\frac{\sqrt p}{V}\right\}.
\end{equation}
The constant is independent of $p$. In particular, for each fixed $p$
the bound is $C_{\mathcal U,p}(\delta_{M,N}+N^{-1}+M^{-d})$.
\end{proposition}
\begin{proof}
Let $A_{ij}=S_{x_i}\cdot S_{x_j}/N$ and $\Sigma=\Sigma_{\beta,J}|_{\mathcal U}$.
They are positive semidefinite with trace $\ell$, and $\Sigma\succeq mI$
by Lemma~\ref{prob:Gram}. For $a_{ij}=\delta_{x_i}-\delta_{x_j}$,
hard length and the unit target diagonal give
\[
 \E(A_{ij}-\Sigma_{ij})
 =\tfrac12D_{a_{ij}}-\tfrac12(K_{a_{ij}}-K_\infty(a_{ij})).
\]
Apply \eqref{ordered:quadratic} to $a_{ij}$, the fixed-path deficit
estimate from Lemma~\ref{prob:norm-saturation}, and the dipole
quadrature estimate from Lemma~\ref{spec:quadrature}. Summing over
the fixed number of entries yields
\begin{equation}
 \|\E A-\Sigma\|_F\le C_{\mathcal U}(\delta+M^{-d}),\qquad
 \E\|A-\Sigma\|_F^2\le C_{\mathcal U}(\delta+N^{-1}+M^{-2d}).
 \label{ordered:gram-moments}
\end{equation}

Put $\bar A=\E A$.
For $a=a_{ij}$, write
$U_a=|\sum_xa_xS_x|^2/N$ and $K_a=Q_M^0(a)/\beta$.
The centred variance satisfies
\[
 \Var(A_{ij})=\tfrac14\Var(U_a)
 \le\tfrac14\E(U_a-K_a)^2
 \le C_{\mathcal U}(\delta+N^{-1})
\]
by \eqref{ordered:quadratic} and the fixed-path bound for $D_a$.
With \eqref{ordered:gram-moments}, the fixed-path estimates therefore
give
\[
 \|\bar A-\Sigma\|_F\le C_{\mathcal U}(\delta_{M,N}+V^{-1}),\qquad
 \E\|A-\bar A\|_F^2\le C_{\mathcal U}(\delta_{M,N}+N^{-1}).
\]
If $\delta$ is below a sufficiently small fixed threshold and $M$ is
large enough, $\bar A\succeq mI/2$.
For an independent standard Gaussian matrix $Z\in\R^{\ell\times p}$,
Lemma~\ref{sector:mixture} applied with lower eigenvalue bound $m/2$
then gives
\[
 W_1(\mathcal L(A^{1/2}Z),\mathcal L(\Sigma^{1/2}Z))
 \le C_{\ell,m}\{p\,\E\|A-\bar A\|_F^2+\sqrt p\,\|\bar A-\Sigma\|_F\}.
\]

Finally use the orbit kernel in Proposition~\ref{prob:Haar}:
$XE_p\overset{\rm law}=A^{1/2}Q_N$, with
$Q_N=\sqrt N PHE_p$ independent of $A$.
Theorem~4.5 of Chatterjee and Meckes \cite[p.~273]{CM},
applied to the $\ell p$ matrices $\sqrt N E_{ji}$, gives
\begin{equation}\label{ordered:haar-block}
 W_1(\mathcal L(Q_N),\mathcal L(Z))
 \le\frac{\sqrt2\,\ell p}{N-1}.
\end{equation}
Their Hilbert--Schmidt inner products are $N\delta_{ii'}\delta_{jj'}$,
as required by that theorem; the Euclidean norm of the vectorized
block is the Frobenius norm used here. The theorem does not require
$\ell p\le N$, so this bound covers every $p\le N$.
Couple $Q_N,Z$ independently of $A$ and use
$\|A^{1/2}\|_{\rm op}\le\sqrt\ell$.
Combining this Haar error with the mixture bound and the two
fixed-path estimates proves \eqref{main:ordered-modular} when
$\delta$ is below the threshold.
If $\delta$ is at or above this threshold, $W_1\le2\sqrt{\ell p}$ is
bounded by $Cp\delta$, so the same estimate holds.
This also covers a singular $\bar A$. For fixed $p$,
\eqref{main:ordered-modular} is the bound
$C_{\mathcal U,p}(\delta_{M,N}+N^{-1}+M^{-d})$, since $V^{-1}=M^{-d}$.
\end{proof}

\subsection{Spin dimension larger than volume}\label{ordered:sec-large}
This subsection proves the following estimate.
\begin{proposition}[Ordered estimate for spin dimension larger than volume]
\label{main:sector}
Fix $\kappa>0$. Under \eqref{main:compact} and
$\beta-\beta_{c,J}\ge\kappa$, there are $C_{\mathcal U,\kappa}$ and
$M_0$ such that, for every integer $M\ge M_0$, $V=M^d$,
$N>V+1$ and $1\le p\le N$,
\[
 W_1(\nu_{M,N,\beta,J}^{\mathcal U,p},\gamma_{\beta,J}^{\mathcal U,p})
 \le C_{\mathcal U,\kappa}
       \left\{\frac{\sqrt p}{V}+\frac{p}{N-V-1}\right\}.
\]
The constant is independent of $p,M,N,\beta,J$ in these ranges.
If $N\ge2(V+1)$, the bound is
$C_{\mathcal U,\kappa}(\sqrt p/V+p/N)$.
For a window containing a translated nearest-neighbour star, if
$N\ge2(V+1)$ and $V\sqrt p\le N$, then
\[
 c\sqrt p/V\le
 W_1(\nu_{M,N,\beta,J}^{\mathcal U,p},\gamma_{\beta,J}^{\mathcal U,p})
 \le C_{\mathcal U,\kappa}\sqrt p/V.
\]
For star-containing windows, the total error is of order $V^{-1}$
for each fixed $p$
when $N\ge2(V+1)$.
\end{proposition}
The density of the
full Gram matrix is a standard matrix-distribution formula; for this
spin model see Tang, Khoo and Ying \cite[Theorem~4]{TangKhooYing}. In their notation, $n=N$, $k=V$,
the unit spin is $S_x/\sqrt N$, and the symmetric interaction matrix
has entries $J_{xy}/2$ on neighbouring pairs, so its double sum counts
each edge once. Combining this density with a Brascamp--Lieb-type variance bound
\cite{BrascampLieb1976} in the form of
\cite[arXiv v2, Theorem~1.1]{MP}
gives the variance bounds below without a factor depending on the volume.

We keep \(\overline S,R,t_x,K_0,r_0\) from
Section~\ref{ordered:sec-mag} and \(h_x,e_x\) from
Section~\ref{ordered:sec-ward}, and write
\(\delta=\delta_{M,N}=(1-V^{-1})/\beta-\E e_x\ge0\). By
\eqref{ordered:gap}, \(r_0\ge r_*:=\kappa/(2B)>0\) for \(M\ge M_0\).
Set $a_{\rm cut}=r_0/4$. The good-event longitudinal energy is
\[
 \G=\frac1V\sum_{\{x,y\}}J_{xy}\E\left[
 \mathbf1_{\{t_x,t_y\ge a_{\rm cut}\}}
          \frac{(t_x-t_y)^2}{t_xt_y}\right].
\]
The sum is over unoriented edges, each once.

\begin{lemma}\label{sector:gram}
If $N>V+1$, then
\[
 \Var(t_x-t_y)\le\frac{C}{N-V-1}
\]
for all sites $x,y$, with an absolute constant. Moreover,
$\Var(A_{xy})\le C/(N-V-1)$ for $A_{xy}=S_x\cdot S_y/N$.
\end{lemma}
\begin{proof}
Under the product sphere prior, the full Gram matrix
$A_{xy}=S_x\cdot S_y/N$ has density
$c_{N,V}(\det A)^\alpha$ on the open convex set
$\mathcal A=\{A\succ0:\operatorname{diag}A=\mathbf1\}$, with
$\alpha=(N-V-1)/2$, relative to off-diagonal Lebesgue measure. This is also the
density in Tang, Khoo and Ying \cite[Theorem~4, Eq.~(2.2)]{TangKhooYing}
at $\beta=0$.
Inductively, suppose $v-1$ independent unit
sphere vectors have Gram matrix $A_{v-1}$. The coordinates $z$ of the
next vector on their orthonormal span have density proportional to
$(1-|z|^2)^{(N-v-1)/2}$. The change of variables from $z$ to its inner
products $u$ with the preceding vectors has Jacobian
$(\det A_{v-1})^{1/2}$. The Schur complement gives
\[
 \det A_v=\det A_{v-1}
              (1-u^{\mathsf T}A_{v-1}^{-1}u).
\]
After including the reciprocal Jacobian and the inductive density
$(\det A_{v-1})^{(N-v)/2}$, every power of $\det A_{v-1}$ cancels,
leaving $(\det A_v)^{(N-v-1)/2}$.
Starting with $v=1$ proves the formula. The full-rank assertion holds
almost surely for $N\ge V$.

The Gibbs tilt is affine in $A$. Thus its negative log density is,
up to a constant,
\[
 f(A)=-\alpha\log\det A-\beta N\sum_{\{z,w\}}J_{zw}A_{zw}
       \quad(A\in\mathcal A),
\]
and is $+\infty$ outside $\mathcal A$. For symmetric tangent matrices
$H$ with zero diagonal,
\[
 D^2f(A)[H,H]=\alpha\tr(A^{-1}HA^{-1}H)>0\quad(H\ne0).
\]
Use Frobenius-orthonormal coordinates on this tangent space. Changing
from off-diagonal coordinates affects only the density's constant.
For any symmetric $L$, the inverse-Hessian quadratic form of the
restricted gradient of $\tr(LA)$ is bounded by
\[
 \sup_{\operatorname{diag}H=0}
 \{2\tr(LH)-\alpha\tr(A^{-1}HA^{-1}H)\}
 \le\frac{\tr(ALAL)}{\alpha}.
\]
The last inequality enlarges the supremum to all symmetric $H$ and
completes the square.

Take $b=\delta_x-\delta_y$, $\mathbf1=(1,\ldots,1)^{\mathsf T}$ and
$L=(b\mathbf1^{\mathsf T}+\mathbf1b^{\mathsf T})/(2V)$. Then
$\tr(LA)=t_x-t_y=:\Delta$, and, with
$U_b=b^{\mathsf T}Ab=|S_x-S_y|^2/N$,
\[
 \tr(ALAL)=\frac{U_bR+\Delta^2}{2}\le4.
\]
Indeed $R\le1$, $U_b\le4$, and $\Delta^2\le U_bR$.
The log-concave linear-variance bound in
\cite[arXiv v2, Theorem~1.1]{MP} now applies directly: its inverse-Hessian
quantity is at most $4/\alpha$ almost surely, so the event in that
theorem has probability one. This yields
$\Var\Delta\le C/\alpha$. A zero restricted gradient instead gives a
constant observable and zero variance. The theorem allows a convex
potential equal to $+\infty$ outside its open domain, as here.
For $x\ne y$, the functional $A_{xy}$ has
$L=(\delta_x\delta_y^{\mathsf T}+\delta_y\delta_x^{\mathsf T})/2$.
Then $\tr(ALAL)=(1+A_{xy}^2)/2\le1$, so the same inequality gives
the claimed entry variance. Diagonal entries are constant.
\end{proof}

\begin{lemma}\label{sector:ward}
For $M\ge M_0$ and $N\ge4$, the actual Gibbs law satisfies, uniformly in the ordered compact ranges,
\[
 |\delta-\G|\le C_\kappa(N^{-1}+V^{-1}).
\]
\end{lemma}
\begin{proof}
Let $U_x=|S_x-\overline S|^2/N$. Gaussian domination for the zero-sum source
$\delta_x-V^{-1}\mathbf1$ and Gaussian linearization give
\[
 \E e^{zNU_x}\le(1-2zK_0)^{-N/2},\qquad 0<2zK_0<1.
\]
Hard length implies $U_x-K_0=R+r_0-2t_x$. Hence $t_x<a_{\rm cut}$ implies
$U_x-K_0>r_0/2$. By \eqref{ordered:gap}, $K_0\le1$ and $r_0\ge r_*$,
so a sufficiently small fixed $z>0$ in the preceding inequality yields
\begin{equation}\label{sector:bad}
 \mathbb P(t_x<a_{\rm cut})\le C_\kappa e^{-c_\kappa N}.
\end{equation}
This one-site bound is uniform in the volume. Since
$V^{-1}\sum_{\{x,y\}}J_{xy}=j$, it also gives
\[
 \frac1V\sum_{\{x,y\}}J_{xy}
 \mathbb P(t_x<a_{\rm cut}\ \text{or}\ t_y<a_{\rm cut})
 \le2j\sup_z\mathbb P(t_z<a_{\rm cut})
 \le C_\kappa e^{-c_\kappa N}.
\]

Choose a smooth function $\varphi$ on $[-1,1]$ equal to $1/t$ for $t\ge a_{\rm cut}$
and zero for $t\le a_{\rm cut}/2$. Its value and derivative have bounds depending
only on $r_*$. At site $x$ use the tangent field
$\varphi(t_x)P_x\overline S$, where $P_x=I-S_xS_x^{\mathsf T}/N$.
Its divergence is
\[
 (N-1)\varphi(t_x)(V^{-1}-t_x)+\varphi'(t_x)(R-t_x^2).
\]
The logarithmic Gibbs derivative along the field is
$\beta N\varphi(t_x)\{t_xe_x-(L_{J,M}t)_x\}$.
Sphere integration by parts, followed by a volume average, gives
\begin{align*}
 0={}&(N-1)\E[\varphi(t_x)(V^{-1}-t_x)]
       +\E[\varphi'(t_x)(R-t_x^2)]
       +\beta N\E[\varphi(t_x)t_xe_x]\\
 &-\frac{\beta N}{V}\sum_{\{x,y\}}J_{xy}
            \E[(\varphi(t_x)-\varphi(t_y))(t_x-t_y)].
\end{align*}
On good sites $\varphi=1/t$; on good edges the product in the last line
is $-(t_x-t_y)^2/(t_xt_y)$. All bad-site and bad-edge replacements
cost $O_\kappa(Ne^{-c_\kappa N})$, by \eqref{sector:bad},
$|t_x|\le1$, $R\le1$, $0\le e_x\le4j$, and bounded $\varphi,\varphi'$.
Each volume-averaged site has only $2d$ neighbours. Substituting
$\E e_x=(1-V^{-1})/\beta-\delta$ and dividing by $N$ yields
\[
 \beta\delta=\frac2N-\frac1V
 +\frac{N-1}{NV}\E\frac{\mathbf1_{\{t_x\ge a_{\rm cut}\}}}{t_x}
 -\frac1N\E\frac{R\mathbf1_{\{t_x\ge a_{\rm cut}\}}}{t_x^2}
 +\beta\G+O_\kappa(e^{-c_\kappa N}).
\]
The two expectations are bounded by $a_{\rm cut}^{-1}$ and $a_{\rm cut}^{-2}$, and
$\beta$ is bounded away from zero. This proves the lemma.
\end{proof}

\begin{proof}[Proof of Proposition~\ref{main:sector}]
Write $q=N-V-1>0$.
Translation invariance gives $\E(t_x-t_y)=0$.
Lemma~\ref{sector:gram}, $t_xt_y\ge a_{\rm cut}^2$ on good edges, and
$V^{-1}\sum_{\{x,y\}}J_{xy}=j$ imply $\G\le C_\kappa/q$.
Lemma~\ref{sector:ward} and $N^{-1}\le q^{-1}$ give
\[
 \delta_{M,N}\le C_\kappa(q^{-1}+V^{-1}).
\]
Now restrict the full Gram matrix to $\mathcal U$, retaining the
notation $A$, and put $\bar A=\E A$, $\Sigma=\Sigma_{\beta,J}|_{\mathcal U}$.
The fixed-path deficit estimate and dipole quadrature, as in
\eqref{ordered:gram-moments}, give
\[
 \|\bar A-\Sigma\|_F\le C_{\mathcal U,\kappa}(q^{-1}+V^{-1}).
\]
Lemma~\ref{sector:gram} bounds the variance of each entry, so
\[
 \E\|A-\bar A\|_F^2\le C_{\mathcal U}/q.
\]
By Lemma~\ref{prob:Gram}, $\Sigma\succeq mI$ uniformly for a fixed
$m>0$. Choose $M_0$ and a fixed $q_0$ large enough that, if
$q\ge q_0$, the bias bound implies $\bar A\succeq mI/2$.
Lemma~\ref{sector:mixture} then gives a mixture error $Cp/q$ and
a covariance bias $C\sqrt p(q^{-1}+V^{-1})$.
The Haar orbit kernel in Proposition~\ref{prob:Haar} and
\eqref{ordered:haar-block} give an additional error $Cp/N$.
Since $p\ge1$ and $N^{-1}\le q^{-1}$, their sum is the asserted bound.
If $0<q<q_0$, both local laws have second moment $\ell p$;
independent coupling gives $W_1\le2\sqrt{\ell p}$.
This is bounded by $Cp/q$ after enlarging $C$ by a constant
independent of $p$. Taking $M_0$ larger also ensures $V\ge\ell$ and $N\ge4$.

If $N\ge2(V+1)$, then $q\ge N/2$ and the upper bound is
$C(\sqrt p/V+p/N)$. On a star-containing window, the first lower
bound in Proposition~\ref{floor:bound}, namely \(c\sqrt p/V\), holds
throughout this regime. The additional condition $V\sqrt p\le N$
therefore yields a matching two-sided $\sqrt p/V$ order. For fixed
$p$ and $N\ge2(V+1)$, $p/N\le p/(2V)$, which gives the two-sided
$V^{-1}$ order on star-containing windows with constants depending
on $p$.
\end{proof}

For $V>N$, the full Gram law is supported on the
nonconvex rank-at-most-$N$ set. At $N=V$ and $N=V+1$ its density
exists but the determinant exponent is nonpositive.
The positive-Hessian argument applies to none of these ranges, and
the restricted estimate alone gives no improved volume-first or accumulation-state rate.

\subsection{Proof of the ordered estimate}\label{ordered:sec-proof}
\begin{proof}[Proof of Theorem~\ref{main:ordered-rate}]
Proposition~\ref{ordered:projection} gives
\eqref{main:ordered-modular}. A direct substitution of
\eqref{ordered:deficit-rate} would introduce \(p/V\). To retain
\(\sqrt p/V\), split into two ranges.

If \(N<2(V+1)\), then \(N\ge4\) implies \(V>N/4\).
The volume part of the deficit in \eqref{ordered:deficit-rate}
can therefore be absorbed into \(C/N\), and
\eqref{main:ordered-modular} gives
\(C_{\mathcal U,\kappa}\{p\psi_d(N)+\sqrt p/V\}\).
If \(N\ge2(V+1)\), Proposition~\ref{main:sector} gives the sharper
bound \(C_{\mathcal U,\kappa}(p/N+\sqrt p/V)\), since
\(N-V-1\ge N/2\). These ranges cover all \(N,V\). Taking the smaller
of this bound and Theorem~\ref{main:joint}, using \(s_M\le C_\kappa/V\)
(Proposition~\ref{main:mass-orders}), proves the asserted upper bound. At each fixed \(N,p\), local second
moments equal \(\ell p\); uniform integrability passes both estimates
to every periodic accumulation state. For \(d\ge5\),
\(p/N\le\sqrt{p/N}\), which gives the simplified rate.

For the lower bounds, put $\bar\chi_q=\E\chi_q$, where $\chi_q$ is the norm
of a standard $q$-dimensional Gaussian vector. At any one site the spin
is uniform on the sphere of radius $\sqrt N$, by rotation invariance.
The $1$-Lipschitz test given by the norm of its first $p$ coordinates
therefore gives, on every nonempty window,
\[
 W_1\ge \bar\chi_p\left(\frac{\sqrt N}{\bar\chi_N}-1\right).
\]
Here $\bar\chi_N=\sqrt2\,\Gamma((N+1)/2)/\Gamma(N/2)$, and the
gamma-ratio expansion gives $\sqrt N/\bar\chi_N-1=1/(4N)+O(N^{-2})$.
For each fixed $p$ this yields a positive lower bound of order
$N^{-1}$. Strict Jensen
gives $\bar\chi_N<\sqrt N$ for every finite $N$, so the finitely many smaller
admissible $N$ can be included in a lower bound $c_p/N$.
On star-containing windows, Theorem~\ref{main:finite-criterion} also
gives $c_pM^{-d}$; the maximum of these two lower bounds controls
$N^{-1}+M^{-d}$. Every periodic accumulation state has the same
one-site sphere marginal. The same norm test thus proves its stated
$N^{-1}$ lower bound, while the upper bound has already passed to
the limit.
\end{proof}

\section{Discussion}\label{main:discussion}
The local law estimate bounds the error in Gaussian predictions.
For $F$ Lipschitz on the $\ell\times p$ coordinate block in the
Frobenius norm, Theorem~\ref{main:joint} gives
\[
 \bigl|\E_{\nu_{M,N,\beta,J}^{\U,p}} F
       -\E_{\gamma_{\beta,J}^{\U,p}} F\bigr|
 \le C\,\operatorname{Lip}(F)\sqrt p\,(N^{-1/2}+s_M).
\]
For example, take distinct sites $x,y\in\U$ and the two observables
$F_\pm(X)=\cos(t(X_{x,1}\pm X_{y,1}))$ with $t\in\R$.
Both have Lipschitz constant at most $\sqrt2\,|t|$, so with $p=1$
each prediction has error at most $C\sqrt2\,|t|(N^{-1/2}+s_M)$.
Write $q=1-\beta_{c,J}/\beta$ for the variance of the Gaussian
constant mode. Under the Gaussian law, the expectations of $F_\pm$ are
$\exp(-t^2 v_\pm/2)$ with
\[
 v_-=\frac{2}{\beta}\bigl(\beta_{c,J}-G_J(x,y)\bigr),
 \qquad
 v_+=\frac{2}{\beta}\bigl(\beta_{c,J}+G_J(x,y)\bigr)+4q .
\]
The constant mode cancels in the spin difference. In the spin sum
it multiplies the GFF prediction by $\exp(-2t^2q)$. The same
observable error bound in periodic accumulation states has no
$s_M$ term, by Corollary~\ref{main:infinite}.

A fixed number of coordinates of a uniformly oriented vector
of length $\sqrt{Nq}$ converge as $N\to\infty$ to independent
Gaussians of variance $q$ \cite{DF}. In the mean spherical reference the constant mode is
Gaussian with variance $1/(\beta V s_M)$, which tends
to $q$ when $\beta>\beta_{c,J}$ is fixed
(Proposition~\ref{spec:crossover}). The same Gaussian zero-mode
mechanism occurs in the continuum mean spherical model with a
cutoff \cite[Appendix~B]{CrisantiSarracinoZannetti}.
In the thermodynamic limit of the isotropic periodic Berlin--Kac
model below its critical temperature, the independent constant
takes the values $\pm\sqrt q$ with equal
probability \cite[Theorem~1.1]{AK}. It produces the factor
$\cos(2t\sqrt q)$ in the spin sum. For $q>0$ and sufficiently small
nonzero $t$, the sum distinguishes the two limiting laws despite
their equal zero-mode variance. The difference between zero-field
spherical and mean spherical averages below criticality was
already noted in \cite{KacThompson1977}.

For fixed $\mathcal U,p$, every fixed-order local mixed moment
also converges to its Gaussian counterpart. Global $O(N)$ invariance gives
\[
 \E|S_x^\alpha|^{2r}
 =\frac{N^r(2r-1)!!}{\prod_{j=0}^{r-1}(N+2j)}
 \le(2r-1)!!,\qquad r\in\mathbb N.
\]
This identity also holds in every periodic accumulation state.
H\"older's inequality, applied with higher-order moments, gives
uniform integrability of each fixed-degree monomial on the coordinate block.
The Gaussian coordinates have the same moment bounds. Truncation and
Corollary~\ref{main:interchange} therefore give, for every fixed polynomial
$P$ on $\R^{\ell\times p}$,
\[
 \sup_{(\beta,J)\text{ satisfying }\eqref{main:compact}}
 \left|\E_{\nu_{M,N,\beta,J}^{\U,p}}P
       -\E_{\gamma_{\beta,J}^{\U,p}}P\right|\longrightarrow0
 \qquad\text{as }\min(M,N)\to\infty.
\]
The corresponding convergence for periodic accumulation states holds
as $N\to\infty$, uniformly over the same parameter range and all choices
of states. The Gaussian expectations are given by Wick's rule at the
same $(\beta,J)$.

For a fixed number of coordinates, the general error bound separates
the choice of box size from the spin-dimensional approximation.
At criticality in $d=3$, $s_M\asymp M^{-2}$
(Proposition~\ref{spec:crossover}), so $M/N^{1/4}\to\infty$ makes the
finite-volume term smaller than the $N^{-1/2}$ term. For $d\ge5$
at a fixed positive ordered gap, Theorem~\ref{main:ordered-rate}
gives $C_{\U,p,\kappa}\operatorname{Lip}(F)(N^{-1}+M^{-d})$ for fixed
$p$. A volume $M^d\ge N$ keeps the finite-volume term no larger
than the spin-dimensional term. The matching lower bound on
star-containing windows is for the $W_1$ distance, which takes
the supremum over unit Lipschitz observables. The coordinate-growth
criterion is exact, while the matching rate concerns fixed $p$.
The optimal rate as a function of growing $p$ remains open.

\paragraph{Infrared bound and order.}
The full-lattice extension of \eqref{main:target} saturates the
Fr\"ohlich--Simon--Spencer infrared bound \cite{FSS}, which
Proposition~\ref{main:gd} gives in the present normalization. Its
spectral measure has density \(\beta^{-1}\lambda_J(k)^{-1}\) at nonzero
frequencies and an atom of weight \(1-\beta_{c,J}/\beta\) at zero.
The unit diagonal is the sum rule. The massless part carries
weight \(\beta_{c,J}/\beta\) and leaves \(1-\beta_{c,J}/\beta\) for the
zero mode. For \(\beta<\beta_{c,J}\), the massless part alone would
exceed the sum rule. The target \eqref{main:target} is therefore
restricted to \(\beta\ge\beta_{c,J}\).

For \(\beta>\beta_{c,J}\) the same bound forces long-range order at
every fixed \(N\) \cite{FSS}. By
Section~\ref{ordered:sec-mag},
\(\E|\overline S|^2/N\ge1-\beta_{c,J}/\beta-O(M^{2-d})\) for the mean
spin \(\overline S=V^{-1}\sum_xS_x\). This is a bound on the atom
at zero of the spectral measure of the normalized spin covariance
\(\E(S_0\cdot S_x)/N\). Along any periodic accumulation sequence
at fixed \(N\), these covariances converge, so the spectral measures
converge weakly. The closed-set
bound for \(\{0\}\) then gives an atom of weight at least
\(1-\beta_{c,J}/\beta\) in the limit.

The two-dimensional continuum linear sigma model has a soft
constraint and a massive Gaussian limit. Delgadino and Smith obtain
an $N^{-1/2}$ estimate for periodic accumulation states in a
weighted Wasserstein distance with cost averaged over components
\cite[Theorem~3.9]{DelgadinoSmith}. Korzhenkova and Sep\'ulveda
study angular freezing and entropic repulsion in a two-dimensional
vector GFF conditioned by a pointwise lower bound on its norm
\cite{KorzhenkovaSepulveda}. Inverse-size rates for fixed marginals
also occur for continuous mean-field Gibbs measures \cite{Lacker}.
For lattice spins at fixed $N$, low-temperature results include
correlation expansions in states selected by a field
\cite{BalabanOCarroll,GO}, a transverse GFF limit on finite graphs
with a pinned spin \cite[Theorem~2.5]{Han}, and estimates uniform
over boundary conditions \cite[Theorem~1.8 and Proposition~1.10]{GarbanMontgolfier}.

A natural open problem is whether
\(W_1(\nu_{M,N,\beta,J}^{\U,p},\gamma_M^{\U,p})\le C\sqrt{p/N}\)
uniformly in \(M\), that is, whether the volume term disappears when
the spin law is compared with the mean spherical reference \(\gamma_M\)
of Remark~\ref{main:gaussian-reference} instead of the target.
For \(N\ge4\), the \(J\)-weighted sums of nearest-neighbour
difference variances agree to within \(CN^{-1/2}\). Let \(v_i^M\) be the
variance of one coordinate of \(X_0-X_{e_i}\) for
\(X\sim\gamma_M^{\U_0,p}\). Then
\(\sum_iJ_i\{Q_M^0(\delta_0-\delta_{e_i})/\beta-v_i^M\}
=s_M-1/(\beta V)\), with \(Q_M^0\) as defined in Section~\ref{main:model}
(see Appendix~\ref{spec:reference}), while
Lemma~\ref{prob:energy-deficit} and the left secant in the proof
of Proposition~\ref{floor:bound} give
\(|\delta_{M,N}-(s_M-1/(\beta V))|\le CN^{-1/2}\) for \(N\ge4\).
Gaussian domination, however, bounds the spin dipole variances by the
massless form, and no sign is available relative to \(\Sigma_M\).
For \(\beta<\beta_{c,J}\), Proposition~\ref{prob:pressure} and both
pressure secants remain valid. For each \(s>0\), Riemann sums give
\(g_M(s)\to g_\infty(s)=\int_{\mathbb T^d}(\lambda_J(k)+s)^{-1}\,d\mu(k)\),
using the notation of Appendix~\ref{spec:appendix}. The function
\(g_\infty\) decreases continuously from \(\beta_{c,J}\) to zero.
Evaluating at positive shifts on either side of the unique solution
\(s_\infty\) of \(g_\infty(s_\infty)=\beta\) brackets \(s_M\) by
monotonicity, so \(s_M\to s_\infty>0\). The energy deficit of
Lemma~\ref{prob:energy-deficit} then satisfies
\(\delta_{M,N}\ge s_M-1/(\beta V)-CN^{-1/2}\); for fixed \(J\) and
\(0<\beta<\beta_{c,J}\) it is therefore bounded away from zero as
\(M,N\to\infty\), so the saturation argument gives no convergence;
moreover, the target \(\gamma_{\beta,J}\) is not defined there.
The known results in that regime are recalled in
Section~\ref{main:introduction}.
Finally, the bounds may be improvable at criticality and in ordered
dimensions three and four; growing windows and states selected by a
field or by boundary conditions are outside the present estimates.

\appendix
\numberwithin{equation}{section}

\section{Spectral estimates}
\label{spec:appendix}
The quadrature and mass bounds below give the volume term in
Theorem~\ref{main:joint} and Proposition~\ref{main:mass-orders}. The final subsection compares the auxiliary
Gaussian reference with the target, as stated in
Remark~\ref{main:gaussian-reference}.

Fix $d\ge3$ and $0<j_0\le j_1<\infty$. Throughout this section,
$J\in[j_0,j_1]^d$, $j=\sum_iJ_i$, $M\ge4$ is an integer, and
$V=M^d$. Frequencies belong to
$\mathbb T^d=\mathbb R^d/(2\pi\mathbb Z)^d$, with normalized measure
$d\mu(k)=dk/(2\pi)^d$, and
\[
 \mathcal K_M=(2\pi/M)\{0,\ldots,M-1\}^d\pmod {2\pi},\qquad
 \lambda_J(k)=2\sum_{i=1}^dJ_i(1-\cos k_i).
\]
Write $\rho(k)$ for the distance from $k$ to zero on this torus. Constants
in the spectral estimates depend only on $d,j_0,j_1$ and, where present,
a fixed source or fixed local set. In mass and Gaussian comparisons we
also fix $0<b\le B<\infty$ and restrict to
$\beta\in[b,B]$, $\beta\ge\beta_{c,J}$. All stated uniform limits have
these compact parameter ranges.

For a finitely supported real source $a$ with $\sum_xa_x=0$, set
\[
 \begin{gathered}
 \widehat a(k)=\sum_xa_xe^{ik\cdot x},\\
 Q_M^0(a)=\frac1V\sum_{k\ne0}
       \frac{|\widehat a(k)|^2}{\lambda_J(k)},\qquad
 Q_\infty(a)=\int_{\mathbb T^d}
       \frac{|\widehat a(k)|^2}{\lambda_J(k)}\,d\mu(k),
 \end{gathered}
\]
where grid sums always range over $\mathcal K_M$. Also define
\[
 \begin{gathered}
 \beta_{c,J}=\int\lambda_J(k)^{-1}\,d\mu(k),\\
 g_M(s)=\frac1V\sum_k\frac1{\lambda_J(k)+s},\qquad
 g_M^0(s)=\frac1V\sum_{k\ne0}\frac1{\lambda_J(k)+s}.
 \end{gathered}
\]
The additive mass $s_M>0$ is the unique solution of $g_M(s_M)=\beta$:
$g_M$ is strictly decreasing, tends to infinity at zero because of the
zero mode, and tends to zero at infinity. Thus
$G_{J,M,s}=(L_{J,M}+sI)^{-1}$; throughout, $s$ denotes the additive
parameter itself.

\subsection{Smooth Fourier quadrature and the zero-sum gain}

\begin{lemma}[Fourier aliasing]
\label{spec:alias}
Let $L>d$ be an integer and $F\in C^L(\mathbb T^d)$. With normalized
$L^1$ norms,
\begin{equation}
 \begin{aligned}
 \frac1V\sum_{k\in\mathcal K_M}F(k)-\int F\,d\mu
   &=\sum_{q\in\mathbb Z^d\setminus\{0\}}\widehat F(Mq),\\
 \left|\frac1V\sum_kF(k)-\int F\,d\mu\right|
   &\le C_{d,L}M^{-L}\max_i\|\partial_i^LF\|_1,
 \end{aligned}
 \label{spec:alias-bound}
\end{equation}
where $\widehat F(n)=\int F(k)e^{-in\cdot k}\,d\mu(k)$. The series on
the right is absolutely convergent.
\end{lemma}

\begin{proof}
For $n\ne0$, select a coordinate $i$ with
$|n_i|=|n|_\infty$. Integrating by parts $L$ times in that coordinate,
with all boundary terms cancelling by periodicity, gives
\[
 |\widehat F(n)|\le |n|_\infty^{-L}
                  \max_i\|\partial_i^LF\|_1.
\]
The shell $|n|_\infty=m$ has at most $C_dm^{d-1}$ points, so
$\sum_{n\ne0}|n|_\infty^{-L}<\infty$. Hence the Fourier series of
$F$ converges absolutely and uniformly, and its sum is $F$ by
uniqueness of Fourier coefficients of continuous functions.

We may now average that series over the finite grid. The geometric sum
\[
 \frac1V\sum_{m\in\{0,\ldots,M-1\}^d}
          e^{2\pi in\cdot m/M}
      =\mathbf1_{\{n\in M\mathbb Z^d\}}
\]
gives the identity in \eqref{spec:alias-bound}; the coefficient bound
and the convergent shell sum give its estimate.
\end{proof}

\begin{lemma}[Sharp zero-sum and diagonal quadrature]
\label{spec:quadrature}
For each fixed finitely supported real zero-sum source $a$,
\begin{equation}
 |Q_M^0(a)-Q_\infty(a)|\le C_aM^{-d}.
 \label{spec:zero-sum}
\end{equation}
Moreover, with $b_M=g_M^0(0)-\beta_{c,J}$,
\begin{equation}
 |b_M|\le CM^{2-d}.
 \label{spec:diagonal}
\end{equation}
Both statements hold for every integer $M\ge4$, uniformly over the
positive compact coupling range.
\end{lemma}

\begin{proof}
For principal representatives in $[-\pi,\pi]^d$,
\begin{equation}
 \frac{4j_0}{\pi^2}\rho(k)^2\le\lambda_J(k)
                \le j_1\rho(k)^2.
 \label{spec:ellipticity}
\end{equation}
The bounds follow from $1-\cos t\ge2t^2/\pi^2$ and
$1-\cos t\le t^2/2$. For any fixed derivative order, the
trigonometric formula gives
$|\partial^\alpha\lambda_J(k)|\le C_\alpha\rho^{2-|\alpha|}$
when $0<\rho\le1$. The cases of order zero and one use the quadratic
and linear vanishing, respectively; at higher orders the derivatives
are bounded. Differentiating $\lambda_J\lambda_J^{-1}=1$ yields
\[
 \partial^\alpha(\lambda_J^{-1})
 =-\lambda_J^{-1}\sum_{0<\eta\le\alpha}
    \binom\alpha\eta(\partial^\eta\lambda_J)
       \partial^{\alpha-\eta}(\lambda_J^{-1}).
\]
Induction and \eqref{spec:ellipticity} therefore give
\begin{equation}
 |\partial^\alpha(\lambda_J^{-1})(k)|
       \le C_\alpha\rho^{-2-|\alpha|}.
 \label{spec:reciprocal-derivatives}
\end{equation}
Away from zero, all these derivatives are uniformly bounded.

Choose a fixed lattice point $x_*$ and multiply $\widehat a$ by
$e^{-ik\cdot x_*}$, which does not change its absolute square. The
zero-sum property gives
\[
 e^{-ik\cdot x_*}\widehat a(k)
   =\sum_xa_x(e^{ik\cdot(x-x_*)}-1)=O_a(\rho).
\]
All positive-order derivatives of this finite sum are bounded. For
$n(k)=|\widehat a(k)|^2$, the product rule consequently gives
$n=O_a(\rho^2)$, $\partial_i n=O_a(\rho)$, and bounded derivatives
of every higher fixed order. In particular,
$|\partial^\alpha n|\le C_{a,\alpha}\rho^{2-|\alpha|}$ on
$0<\rho\le1$. Together with
\eqref{spec:reciprocal-derivatives}, this proves
\begin{equation}
 f_a(k)=\frac{|\widehat a(k)|^2}{\lambda_J(k)},\qquad
 |\partial^\alpha f_a(k)|\le C_{a,\alpha}\rho^{-|\alpha|}.
 \label{spec:source-derivatives}
\end{equation}
Set $f_a(0)=0$ to implement removal of the zero grid mode.

Fix $\chi\in C_c^\infty(\mathbb R^d)$ with $0\le\chi\le1$,
$\chi=1$ on $|u|\le1$, and $\chi=0$ on $|u|\ge2$. The function
$\chi(Mk)$, defined in the principal cell and then periodically
extended, has support strictly inside that cell. Put
$F_M=(1-\chi(Mk))f_a(k)$ and $F_M(0)=0$. It is smooth and periodic.
Every nonzero grid point has torus distance at least $2\pi/M>2/M$,
so its grid average is exactly $Q_M^0(a)$. Its integral differs from
$Q_\infty(a)$ by at most $C_aM^{-d}$. For an integer $L>d$,
\[
 \max_i\|\partial_i^LF_M\|_1\le C_aM^{L-d}.
\]
Indeed, outside the transition annulus this follows by integrating
$\rho^{-L}$ from $1/M$ to a fixed radius; on the annulus each
Leibniz term is $O_a(M^L)$ and the volume is $O(M^{-d})$. The
derivatives vanish inside the inner ball. Applying
Lemma~\ref{spec:alias} proves \eqref{spec:zero-sum}.

For the diagonal estimate use instead
$F_M=(1-\chi(Mk))\lambda_J(k)^{-1}$. Its grid average is
$g_M^0(0)$, the removed integral is $O(M^{2-d})$, and
\eqref{spec:reciprocal-derivatives} gives
\[
 \max_i\|\partial_i^LF_M\|_1\le CM^{L+2-d}.
\]
The same aliasing estimate proves \eqref{spec:diagonal}. In both
applications the Fourier formula is applied to the smooth cutoff
function, whose derivative norms depend on $M$.
\end{proof}

The order in \eqref{spec:zero-sum} is sharp for this class of fixed
sources. For $b_i=\delta_0-\delta_{e_i}$,
$\sum_iJ_i|\widehat b_i(k)|^2=\lambda_J(k)$, and hence
\begin{equation}
 \begin{aligned}
 \sum_iJ_i\{Q_M^0(b_i)-Q_\infty(b_i)\}&=-M^{-d},\\
 \max_i|Q_M^0(b_i)-Q_\infty(b_i)|
      &\ge\frac{M^{-d}}{j}.
 \end{aligned}
 \label{spec:zero-sum-sharp}
\end{equation}
The individual directional errors may have either sign.

\subsection{The mass equation and the crossover}

Define
\[
 \begin{gathered}
 H_M(s)=\frac1V\sum_{k\ne0}
            \frac1{\lambda_J(k)(\lambda_J(k)+s)},\\
 a_d(M)=\begin{cases}M,&d=3,\\\log M,&d=4,\\1,&d\ge5,\end{cases}
 \qquad t_d(M)=\sqrt{a_d(M)/V},
 \qquad \rho_d(M)=(Va_d(M))^{-1/2}.
 \end{gathered}
\]

\begin{lemma}[Centred-cell bound and resolvent shells]
\label{spec:cell-shell}
For $\beta\ge\beta_{c,J}$,
\begin{equation}
 \frac1{\beta V}\le s_M\le
 \eta_M:=2j\left(1-\frac{\sin(\pi/M)}{\pi/M}\right)
        \le\frac{j\pi^2}{3M^2}.
 \label{spec:cell-mass}
\end{equation}
For each fixed $C_0<\infty$, uniformly for $0\le s\le C_0M^{-2}$,
\begin{equation}
 c_{C_0}a_d(M)\le H_M(s)\le C a_d(M).
 \label{spec:shell-bound}
\end{equation}
\end{lemma}

\begin{proof}
The lower bound in \eqref{spec:cell-mass} follows immediately from
$\beta=g_M(s_M)\ge1/(Vs_M)$. To obtain the upper bound, partition
the Fourier torus into cells centred at the points of $\mathcal K_M$,
each with coordinate half-width $\pi/M$. Put
$c_M=\sin(\pi/M)/(\pi/M)$. The cell average of
$\cos(k_i+u_i)$ is $c_M\cos k_i$, so the average of $\lambda_J$
over the cell centred at $k$ is exactly
\[
 c_M\lambda_J(k)+2j(1-c_M)\le\lambda_J(k)+\eta_M.
\]
Jensen's inequality for the convex function $x\mapsto1/x$ therefore
gives the cell average of $\lambda_J^{-1}$ at least
$(\lambda_J(k)+\eta_M)^{-1}$. At the cell containing zero, apply
Jensen first to $(\lambda_J+\varepsilon)^{-1}$ and then let
$\varepsilon\downarrow0$; the integral is finite since $d>2$ and
\eqref{spec:ellipticity} holds. Summing the cells proves
$\beta_{c,J}\ge g_M(\eta_M)$. Strict decrease of $g_M$, together
with $g_M(s_M)=\beta\ge\beta_{c,J}$, implies $s_M\le\eta_M$.
Finally, $1-\sin u/u\le u^2/6$ yields the last bound in
\eqref{spec:cell-mass}.

Choose centred integer representatives $k=2\pi q/M$ with
$|q_i|\le M/2$. By \eqref{spec:ellipticity} and the bound
$O(n^{d-1})$ for the shell $|q|_\infty=n$,
\[
 H_M(0)\le CM^{4-d}
       \sum_{n=1}^{\lceil M/2\rceil}n^{d-5}
       \le Ca_d(M).
\]
For the reverse bound in $d=3$, the single frequency $q=e_1$ gives
$H_M(0)\ge cM$. In $d=4$, use the full shells
$1\le n\le\lfloor M/4\rfloor$: they have
$(2n+1)^4-(2n-1)^4\ge cn^3$ points, and
$\lambda_J(2\pi q/M)\le Cn^2/M^2$ there. Their contributions sum
to at least $c\log M$, after adjusting the constant for the finitely
many small values of $M$. For $d\ge5$, the global bound
$\lambda_J\le4j$ gives
$H_M(0)\ge(1-V^{-1})/(4j)^2\ge c$. Thus
$H_M(0)\asymp a_d(M)$.

Every nonzero eigenvalue is at least $cM^{-2}$. For
$s\le C_0M^{-2}$, the multiplier
$\lambda_J(k)/(\lambda_J(k)+s)$ lies between a positive constant
depending on $C_0$ and one. This proves \eqref{spec:shell-bound}.
\end{proof}

\begin{proposition}[Uniform critical-to-ordered crossover]
\label{spec:crossover}
With $\tau=\beta-\beta_{c,J}\ge0$, there are uniform positive
constants $c,C$ such that
\begin{equation}
 \frac{c}{V\{\tau+t_d(M)\}}\le s_M\le
 \frac{C}{V\{\tau+t_d(M)\}}.
 \label{spec:crossover-bound}
\end{equation}
At criticality this gives
\begin{equation}
 s_M\asymp\rho_d(M)=\begin{cases}
 M^{-2},&d=3,\\
 M^{-2}(\log M)^{-1/2},&d=4,\\
 M^{-d/2},&d\ge5.
 \end{cases}
 \label{spec:critical-orders}
\end{equation}
If $\tau\ge\kappa>0$, then, uniformly on that parameter set,
\begin{equation}
 \frac1{\beta V}\le s_M\le\frac2{\kappa V}
 \quad(M\ge M_0(\kappa)),\qquad
 Vs_M=\frac1\tau+O_\kappa(M^{2-d}).
 \label{spec:ordered-asymptotic}
\end{equation}
\end{proposition}

\begin{proof}
The exact resolvent identity
$g_M^0(s)=g_M^0(0)-sH_M(s)$ rewrites the mass equation as
\begin{equation}
 \begin{aligned}
 \frac1{Vs_M}&=\tau-b_M+s_MH_M(s_M),\\
 1&=Vs_M(\tau-b_M)+Vs_M^2H_M(s_M).
 \end{aligned}
 \label{spec:mass-identity}
\end{equation}
At $\tau=0$, the following calculation uses only the bounds on $J$,
even when $\beta_{c,J}$ lies outside $[b,B]$. Put $z=s_M/\rho_d(M)$. By
Lemma~\ref{spec:cell-shell}, $H_M(s_M)\asymp a_d(M)$; by
\eqref{spec:diagonal},
$|Vs_Mb_M|\le CzM^2\rho_d(M)$. Here
\[
 M^2\rho_d(M)=\begin{cases}
 1,&d=3,\\(\log M)^{-1/2},&d=4,\\M^{2-d/2},&d\ge5,
 \end{cases}
\]
is uniformly bounded. The critical mass identity consequently implies
\[
 cz^2-Cz\le1\le Cz^2+Cz.
\]
The first inequality bounds $z$ above by the positive root of a fixed
quadratic; the second bounds it below by a positive root. This proves
\eqref{spec:critical-orders}.

Write $t=t_d(M)$, so $\rho_d(M)=1/(Vt)$. In all dimensions,
\eqref{spec:diagonal} implies $|b_M|\le C_b t$. Since $s_M$
decreases as $\beta$ increases, the critical estimate gives
$s_M\le C\rho_d(M)$ throughout $\tau\ge0$. Let
$L_0=\max(1,2C_b)$. If $\tau\ge L_0t$, then
$\tau-b_M\ge\tau/2$, so \eqref{spec:mass-identity} gives
\[
 s_M\le\frac2{V\tau}\le\frac4{V(\tau+t)}.
\]
If $\tau<L_0t$, the critical upper bound gives
$s_M\le C/(Vt)\le C(L_0+1)/[V(\tau+t)]$. This proves the
upper bound in \eqref{spec:crossover-bound}.

For the lower bound, set $u=Vs_M(\tau+t)$. Using
$H_M(s_M)\le Ca_d(M)$ in the second identity of
\eqref{spec:mass-identity} gives
\[
 1\le u\frac{\tau+C_bt}{\tau+t}
      +Cu^2\frac{a_d(M)/V}{(\tau+t)^2}
    \le C'(u+u^2).
\]
Thus $u$ is bounded below by a uniform positive constant. Both sides
of \eqref{spec:crossover-bound} hold also when $\tau$ varies with $M$.

Finally, if $\tau\ge\kappa$, choose $M_0$ so that
$|b_M|\le\kappa/2$. The first mass identity implies
$s_M\le2/(\kappa V)$, and the zero mode gives the stated lower
bound. The shell estimate now yields
\[
 0\le s_MH_M(s_M)\le C_\kappa V^{-1}a_d(M)
                         \le C_\kappa M^{2-d}.
\]
Consequently $1/(Vs_M)=\tau+O_\kappa(M^{2-d})$, with the
denominator bounded below by $\kappa/2$. Taking its reciprocal proves
\eqref{spec:ordered-asymptotic}.
\end{proof}

The finite-size behaviour of the spherical-model mass is classical;
see \cite{BrankovDanchevTonchev2000}. Related finite-size analyses of
spherical, large-$N$ and $\varphi^4$ models include
\cite{CGGP,Chamati,ChenDohm,SinghPathria1987,SinghPathria}.

\subsection{The auxiliary Gaussian reference}\label{spec:reference}
Let $\mathcal U$ be a fixed set of distinct lattice points and
$\ell=|\mathcal U|$, with $M$ large enough that it embeds injectively
in the torus. Define the covariance matrices restricted to this set by
\[
 \Sigma_M(x,y)=\beta^{-1}G_{J,M,s_M}(x,y),\qquad
 \Sigma_\infty(x,y)=\beta^{-1}G_J(x,y)+1-\beta_{c,J}/\beta.
\]
Here \(\Sigma_\infty=\Sigma_{\beta,J}\) is the target covariance
\eqref{main:target}, and both matrices have diagonal one. As in
Remark~\ref{main:gaussian-reference}, denote by
$\gamma_M^{\mathcal U,p}$ and $\gamma_{\beta,J}^{\mathcal U,p}$
the laws of $p$ independent centred Gaussian columns with these
covariances. The metric below is $W_1$ for the Euclidean norm on
$\mathbb R^{\ell\times p}$.
We now prove the bounds stated in Remark~\ref{main:gaussian-reference}.
\begin{proof}
For a fixed zero-sum source let
$Q_{M,s}(a)=V^{-1}\sum_{k\ne0}|\widehat a(k)|^2/(\lambda_J(k)+s)$.
By \eqref{spec:ellipticity} and the first-order vanishing of
$\widehat a$, $|\widehat a(k)|^2\le C_a\lambda_J(k)$ on the torus.
Consequently
\[
 0\le Q_M^0(a)-Q_{M,s_M}(a)
  =\frac{s_M}{V}\sum_{k\ne0}
       \frac{|\widehat a(k)|^2}{\lambda_J(k)(\lambda_J(k)+s_M)}
  \le C_as_Mg_M^0(0)\le C_as_M,
\]
where \eqref{spec:diagonal} bounds $g_M^0(0)$ uniformly. For each
dipole $a=\delta_x-\delta_y$, its variance under $\Sigma_M$ is
$Q_{M,s_M}(a)/\beta$, and under $\Sigma_\infty$ it is
$Q_\infty(a)/\beta$ because the constant mode cancels. The diagonal
entries are one, so Lemma~\ref{spec:quadrature} gives
\[
 \|\Sigma_M-\Sigma_\infty\|_F\le C_{\mathcal U}(s_M+M^{-d})
                           \le C_{\mathcal U}s_M.
\]
The last inequality uses $M^{-d}\le\beta s_M\le Bs_M$ from
\eqref{spec:cell-mass}.

By \eqref{prob:target-positive}, $\Sigma_\infty\succeq mI$ with
$m=(4\beta j)^{-1}$.
The square-root estimate \eqref{prob:sqrt-Lipschitz}, applied with
$A=\Sigma_M$ and $\Sigma=\Sigma_\infty$, gives
\[
 \|\Sigma_M^{1/2}-\Sigma_\infty^{1/2}\|_F
 \le m^{-1/2}\|\Sigma_M-\Sigma_\infty\|_F.
\]
Set $E_\Sigma=\Sigma_M^{1/2}-\Sigma_\infty^{1/2}$. Coupling the two Gaussian blocks
by the same $\ell\times p$ standard Gaussian matrix $Z$ gives
\[
 W_1\le \E\|E_\Sigma Z\|_F\le\sqrt p\,\|E_\Sigma\|_F,
\]
which proves \eqref{spec:gaussian-upper}.

For the lower bound let $v_i^M,v_i^\infty$ be the single-column
variances of $X_0-X_{e_i}$ under the two laws. Since
$\sum_iJ_i|\widehat b_i|^2=\lambda_J$, the mass equation gives
the exact identities
\[
 \sum_iJ_iv_i^M=\frac1{\beta V}\sum_k
                 \frac{\lambda_J(k)}{\lambda_J(k)+s_M}
              =\frac1\beta-s_M,
 \qquad \sum_iJ_iv_i^\infty=\frac1\beta.
\]
Thus
\begin{equation}
 \sum_iJ_i(v_i^\infty-v_i^M)=s_M,
 \qquad v_i^\infty-v_i^M\ge s_M/j\quad\hbox{for some }i.
 \label{spec:weighted-variance}
\end{equation}
The second assertion uses only the weighted average; individual
differences may have either sign. The selected direction may depend on
$M,\beta,J$, and the star contains all possible directions.

Together with \(\sum_iJ_iQ_M^0(b_i)=1-V^{-1}\) from the proof of
Lemma~\ref{prob:energy-deficit}, the first identity also gives the
Gaussian deficit
\(\sum_iJ_i\{Q_M^0(b_i)/\beta-v_i^M\}=s_M-1/(\beta V)\) used in
Sections~\ref{main:proof} and~\ref{main:discussion}.

For centred isotropic Gaussian laws in $\mathbb R^p$, let $G$ be a
standard Gaussian vector and write $\chi_p=|G|$. Then
\[
 W_1(\mathcal N(0,vI_p),\mathcal N(0,wI_p))
       =\E\chi_p\,|\sqrt v-\sqrt w|.
\]
Coupling the two laws with the same $G$ gives the upper bound, and
the $1$-Lipschitz test $|x|$ gives equality. Pointwise
$|G|\ge p^{-1/2}\sum_{\alpha=1}^p|G_\alpha|$, so
\[
 \E\chi_p\ge\sqrt{2/\pi}\,\sqrt p.
\]
The two laws of the dipole vector
$(X_0^\alpha-X_{e_i}^\alpha)_{\alpha=1}^p$ have covariances
$v_i^MI_p$ and $v_i^\infty I_p$, since their Gaussian columns are
independent. Both dipole variances are at most four because the site
variances are one. Their standard deviations therefore sum to at most
four, and \eqref{spec:weighted-variance} yields vector distance at
least $\sqrt{2/\pi}\,\sqrt p\,s_M/(4j)$. The map from the entire
block to this dipole vector has Lipschitz constant $\sqrt2$.
Contraction of $W_1$ under that map proves
\eqref{spec:gaussian-lower}. Projection proves the assertion for a
larger local set.
\end{proof}

\section*{Statements and Declarations}
\paragraph{Acknowledgements.}
We acknowledge Gewu Intelligence Lab for providing the collaborative
research environment in which this project was developed. Large language
models (OpenAI Codex and Anthropic Claude) assisted with literature
retrieval, proof development and checking, and editing. The authors
checked the resulting derivations and take full responsibility for the
scientific content and conclusions of this paper.
\paragraph{Competing interests.} The authors declare no competing interests.
\paragraph{Data availability.}
No datasets were generated or analysed in this study.

\end{document}